\documentclass{article}
\usepackage{graphicx,float}
\usepackage{amssymb,amsfonts,amsmath,amsthm}
\usepackage{multicol}
\usepackage{hyperref}
\newtheoremstyle{problemstyle}  
        { }                                               
        { }                                               
        {\sf}  
        {15pt}                                                  
        {\bfseries\itshape}                 
        {\normalfont\bfseries:}         
        {.5em}                                          
        {}            
\theoremstyle{problemstyle}
\newtheorem{problem}{Problem}[section] 

\hypersetup{%
linkcolor=black,%
citecolor=black,%
linkbordercolor={1 1 1},%
citebordercolor={1 1 1}%
}%

\newcommand{\figpend}{%
\begin{figure}[htbp]
\begin{center}
\includegraphics[scale=0.51,clip]{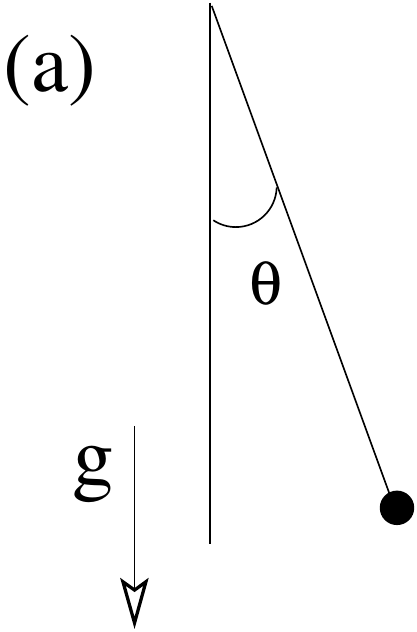}\qquad \qquad 
\includegraphics[scale=0.651,clip]{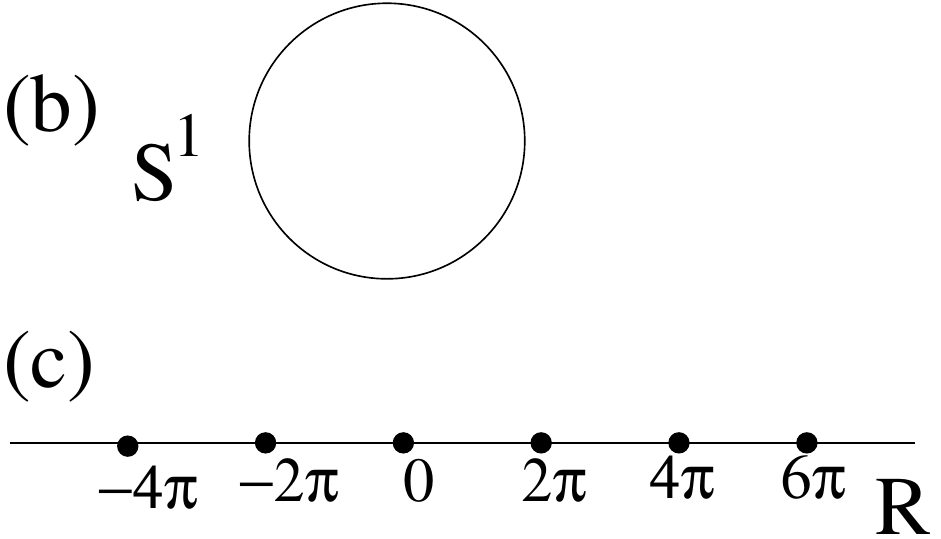}
\end{center}
\label{fig:11}
\caption{Planar pendulum: (a) A bob of mass $m (=1)$ is suspended by a
  massless rigid rod of length $L (=1)$ in a uniform gravitational
  field with $g (=1)$ as the acceleration due to gravity.  The
  generalized coordinate $q$ is angle $\theta$.  The configuration
  space for $\theta$ is (b) a circle $S^1$ or, (c) the real axis with
  equivalent points $x=x+2n\pi, n\in\mathbb{Z}$. Here ${\mathbb{Z}}$
  represents the set of all integers, positive, negative, 0. We may
  choose $q=x$ to be the length of the arc along the circular contour.
  A simple harmonic oscillator corresponds to $q=x\in R$ without any
  equivalent point.  }\label{fig:pend1}
\end{figure}
}%

\newcommand{\figphspace}{%
\begin{figure}[htbp]
\begin{center}
\includegraphics[scale=0.51,clip]{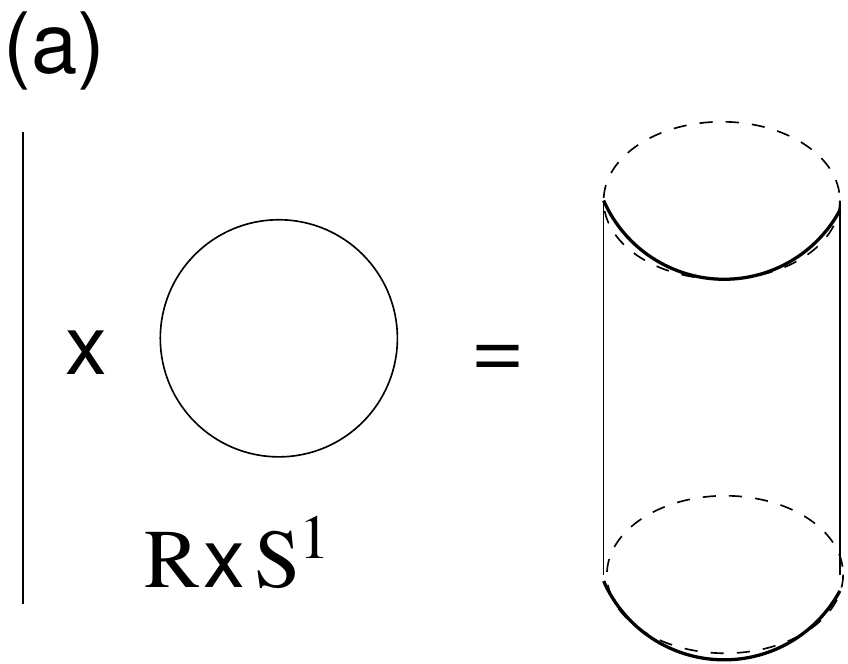}\qquad \qquad 
\includegraphics[scale=0.651,clip]{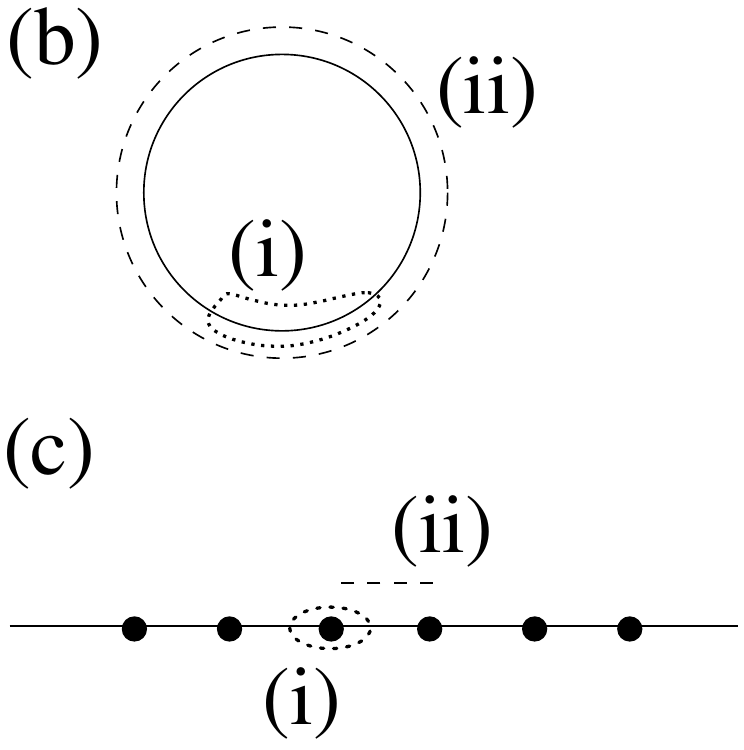}
\end{center}
\label{fig:12}
\caption{(a)The direct product phase space is a cylinder. (b) Type-1
  and type-2 trajectories are marked (i) and (ii) in (b) and (c).  The
  trajectory going completely around the circle in (b) maps to an open
  line in (c).  Consequently any closed loop on ${\mathbb{R}}$ can be
  shrunk to a point.  The real line is the {\it universal cover} of
  $S^1$.  }\label{fig:pend2}
\end{figure}
}%

\newcommand{\figcylorb}{%
\begin{figure}[htbp]
\begin{center}
{\bf \large (a)} \includegraphics[scale=0.41,clip]{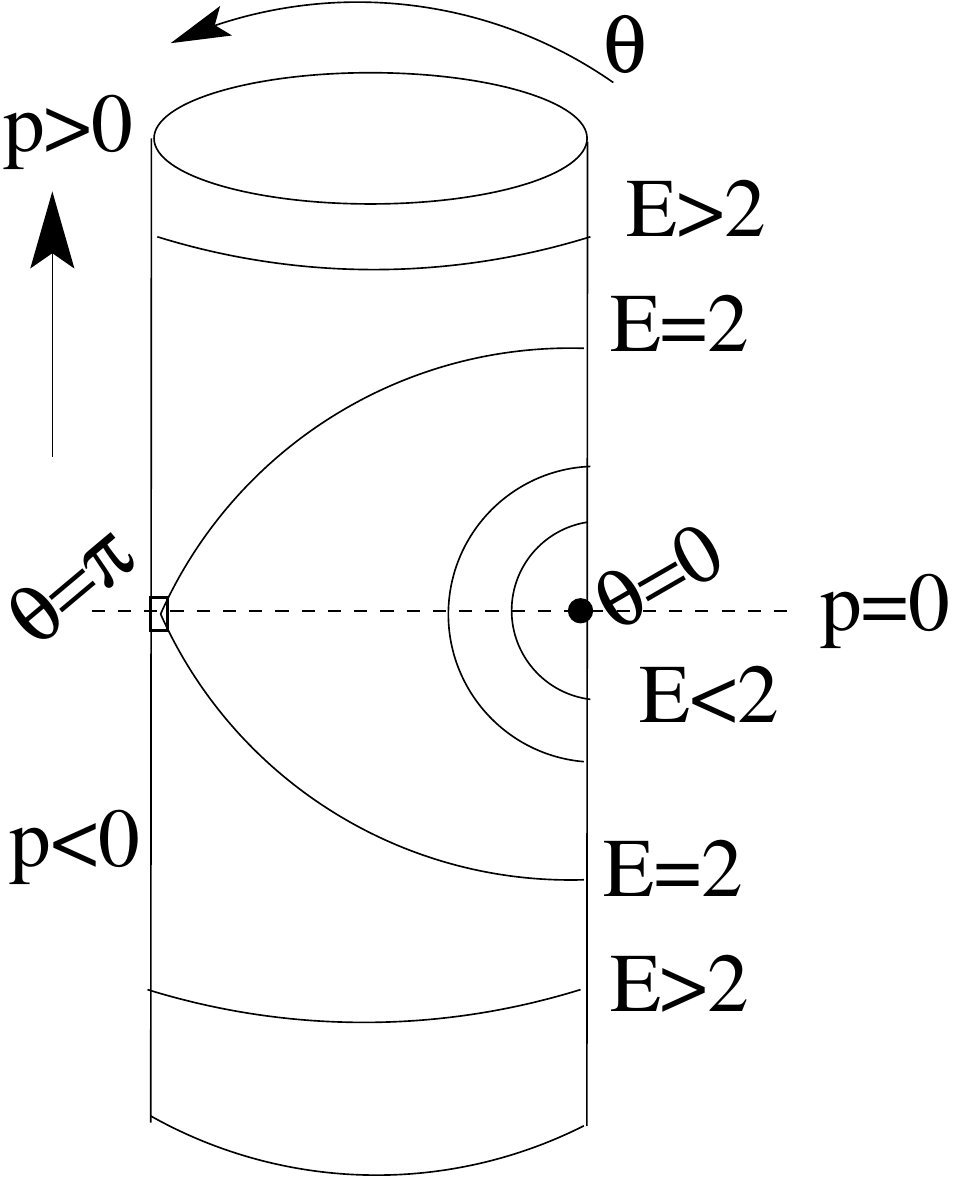}\qquad \qquad
{\bf\large (b)} \includegraphics[scale=0.851,clip]{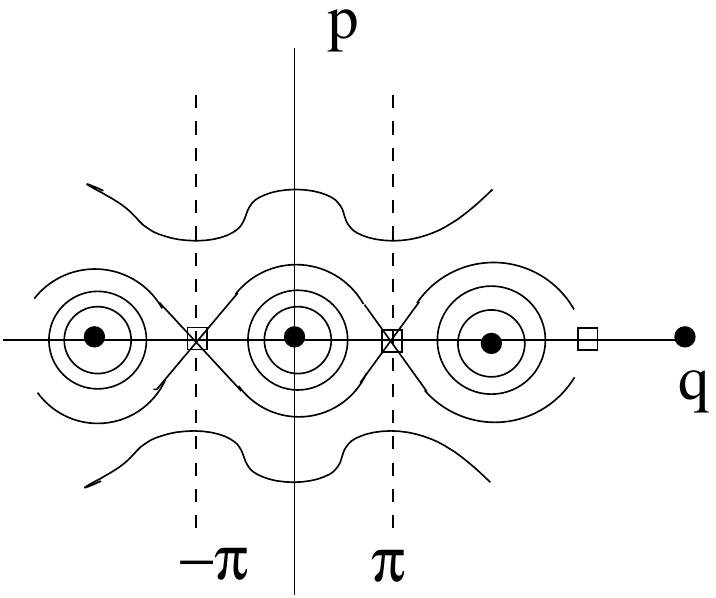} 

{\bf \large (c)} \includegraphics[scale=0.51,clip]{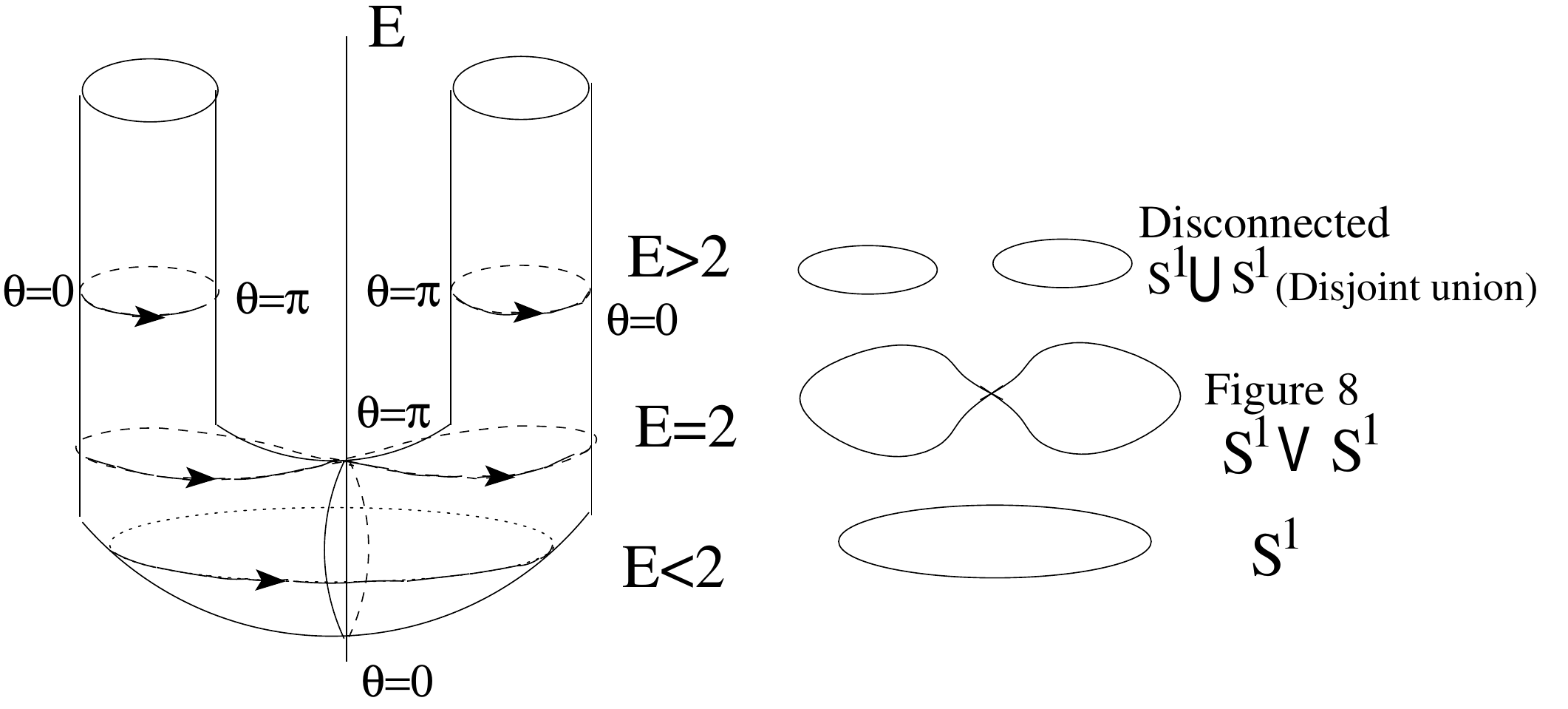} \qquad
\end{center}
\caption{Trajectories on (a) the cylindrical phase space, and (b) the
  extended space ${\mathbb{R}}^2$. The closed orbits around $\theta=0$
  in (a) are equivalent to the closed orbits around the stable points
  $q=2n\pi$ denoted by filled dots in (b) for $E<2$.  In contrast, the
  closed orbits in (a) encircling the cylinder for $E>2$ correspond to
  the open ones in (b).  The critical trajectory connects the unstable
  points $q=(2n+1)\pi$ represented by unfilled squares in (b) but it
  just encircles the cylinder in (a) with a point of contact.  The
  U-tube space when $E$ is used as an axis is shown in (c).  The
  actual trajectories are the $E=$const plane intersections of the
  U-tube.  On the right, three different types of intersections:
  closed for $E<2$ (topologically equivalent to a circle $S^1$),
  figure 8 for $E=2$ (two circles with  one common point, called
  the wedge sum $S^1\vee S^1$), and two disconnected
  closed pieces for $E>2$ (disjoint union of two circles, $S^1\cup
  S^1$ but with $S^1\cap S^1=\emptyset$ ).}\label{fig:pend3}
\end{figure}
}%

\newcommand{\figinfb}{%
\begin{figure}[htbp]
\begin{center}
\includegraphics[scale=0.51,clip]{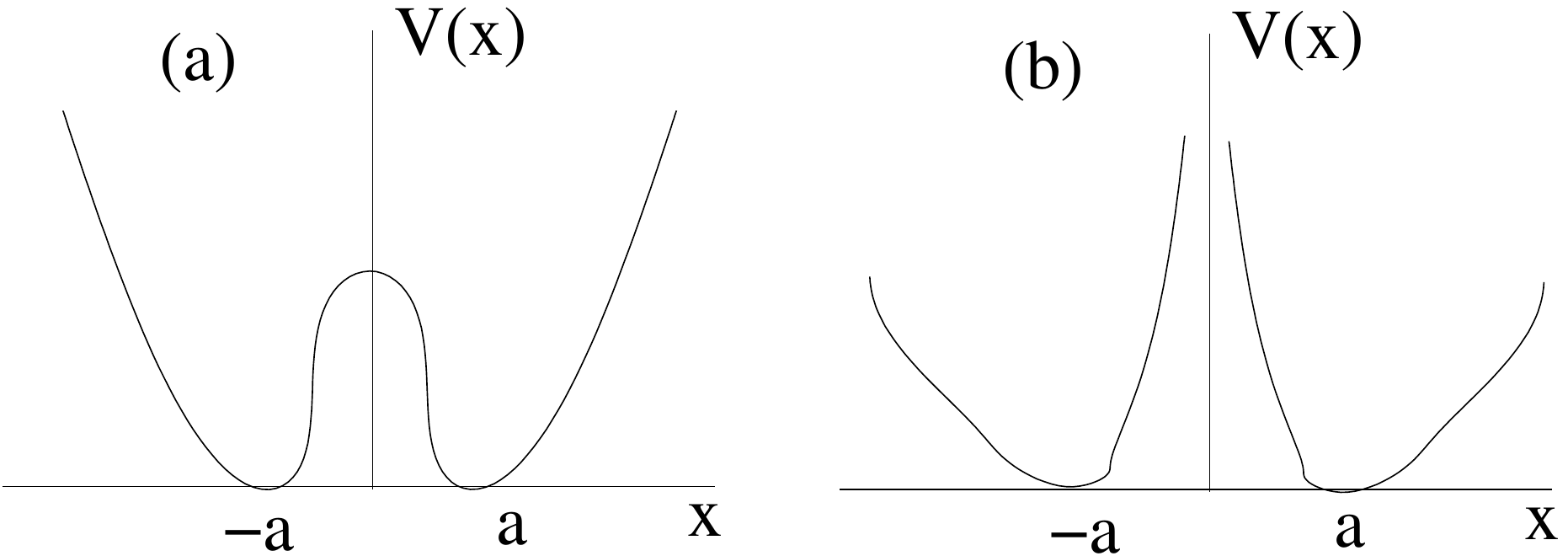}\qquad \qquad 
\end{center}
\label{fig:14}
\caption{(a) A  double well potential. (b) A double well potential but
  with an infinite barrier inbetween.  The barrier cannot be crossed.
 }\label{fig:infb}
\end{figure}
}%

\newcommand{\figtwoc}{%
\begin{figure}[htbp]
  \begin{center}
\includegraphics[scale=0.5,clip]{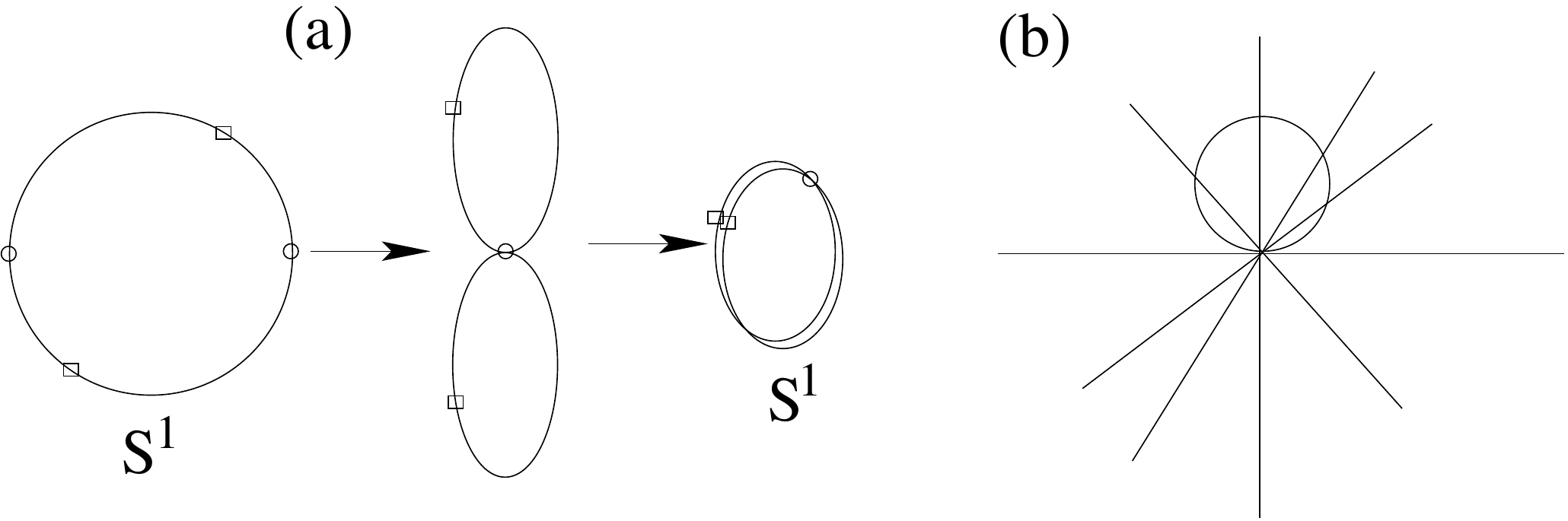}    
  \end{center}
\label{fig:15}
\caption{(a) Twist a circle which brings two antipodal points together.
Then fold the two circles so that again antipodal points are on top of
eachother. (b) The space of all lines through origin. Any point on a
line, except the origin, are equivalent to all others on the same
line. This space is $S^1$.}\label{fig:twoc}
\end{figure}
}%

\newcommand{\figbasis}{%
\begin{figure}[H]
\begin{minipage}[c]{0.4\textwidth}
\begin{center}
\includegraphics[width=0.9\linewidth,angle=0,clip]{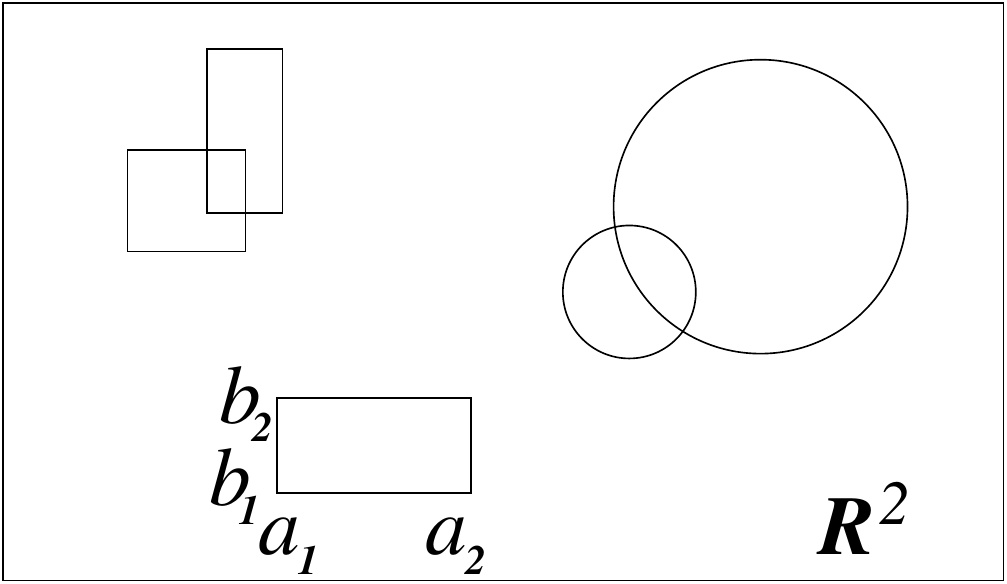}
\end{center}
\end{minipage}\hfill
\begin{minipage}[c]{0.55\textwidth}
\label{fig:16}
\caption{Basis for   $R^2$ as a topological space: open rectangles
or disks.
}\label{fig:r2}
\end{minipage}
\end{figure}
}%
\begin{document}


\title{Use of Topology in physical problems}
\author{Somendra M Bhattacharjee\\
{Institute of Physics, Bhubaneswar 751005, India}\\
and Homi Bhabha National Institute\\
 Training School Complex, Anushakti Nagar, Mumbai 400085, India\\
{email:somen\string@iopb.res.in}
}
\maketitle

\hrule 
\vspace{6pt}

Some of the basic concepts of topology are explored through known
physics problems.  This helps us in two ways, one, in motivating the
definitions and the concepts, and two, in showing that topological
analysis leads to a clearer understanding of the problem.  The
problems discussed are taken from classical mechanics, quantum
mechanics, statistical mechanics, solid state physics, and biology
(DNA), to emphasize some unity in diverse areas of physics.

It is the real Euclidean space, ${\mathbb{R}}^d$, with which we are
most familiar.  Intuitions can therefore be sharpened by appealing to
the relevant features of this known space, and by using these as simplest
examples to illustrate the abstract topological concepts.  This is
what is done in this chapter.  \vspace{6pt}

\hrule 

\begin{multicols}{2}
\tableofcontents
\end{multicols}
\hrule 
\vspace{6pt}

\section{The Not-so-simple  Pendulum}
An ideal pendulum is our first example.  It is not necessarily  a simple harmonic
oscillator (SHO),  though the small amplitude motion can be
well approximated by a linear oscillator.  This difference is
important for dynamics, and a topological analysis brings that out.

The planar motion of a pendulum in the earth's gravitational field is
described by a generalized coordinate $q$ where $q$ is the angle
$\theta$ as in Fig.  \ref{fig:pend1}.

\subsection{Mechanics}
\label{sec:mechanics}
The equation of motion (with all constants set to $1$) can be written
as a second order equation or two first order equations involving the
momentum $p$ as
\begin{equation}
  \label{eq:1}
 \ddot{q}
+ \sin  q =0,\ {\rm or}\ 
\left\{ \begin{array}{ll}
  \dot{q}&=p, \\ 
\dot{p}&=-\sin q,
\end{array}\right.
\end{equation}
where a dot represents a time derivative, and the conserved energy as
 \begin{equation}
   \label{eq:3}
   E=\frac{1}{2} \; \dot{q}^2 + (1-\cos q).
\end{equation}

\figpend

Let us make a list of some of the relevant results known from
mechanics.
\begin{enumerate}
\item The potential energy has minima at $ q=2n\pi$, and maxima at $
  q=(2n+1)\pi$, $n=0,\pm1,...$, i.e., $n\in {\mathbb{Z}}$.  These represent the
  {\it stable} and the  {\it unstable} equilibrium points.
\item The stable motion for small energies, $E<2$, are 
  oscillations around the minimum energy point $ q=0$. Let's call
  these {\it type-O} motion.
\item For larger energies, $E>2$, the motion consists of rotations in
  the vertical plane, clockwise or anticlockwise. These are 
  {\it type-R} motion.
\item There is a very special critical one that separates the above
  two types, viz., the case with $E=2$, when $E$ is equal to the
  potential energy at the topmost position ($
  q=\pi$).  Let's call it {\it type-C}.\\
  The strangeness of the critical one is its infinite time
  period\footnote{This can be seen by integrating Eq. \ref{eq:3} for
    the time taken to go from $q=0$ to $q=\pi$ as $\int_0^{\pi}
    \sqrt{\sec (q/2)\; }\;\, dq \to \infty$ (from $q\to\pi)$.}.  Since
  most of the time is spent near the top, it looks like an inverted
  pendulum at the unstable equilibrium point.
\item For all problems of classical mechanics or statistical mechanics,
there are two spaces to deal with, the {\it configuration space} for
the set of values taken by the degrees of freedom, and the {\it phase
  space}, where the configuration space is augmented by the set of
values of the momenta.  {\it What sort of ``spaces'' are these}?

\item That there are three different classes of orbits cannot be
  overemphasized.  The equation of motion is time reversible under
  $t\to -t, q\to q, p\to -p$.  This time reversibility is respected by
  the type-O motion, but not by type-R because a right circular motion
  would go over to a left one.  For type-R, the symmetry
  is explicitly broken by the initial conditions, which, however, do
  not play any crucial role for type-O.

\item As a coupled first order equations, Eq. (\ref{eq:1}), has fixed
  points at $q=n\pi,p=0$, which are {\it centres} for {\it even $n$}
  but {\it saddle points} for {\it odd} $n$. 
\end{enumerate}

\subsection{Topological analysis: Teaser }
\label{sec:topological-analysis}
A topological analysis of the motion would be based on possible
continuous deformations of one solution or trajectory to the other,
without involving  any explicit  solution of  Eq.  \ref{eq:1}.

Why deform?  This is tantamount to asking whether there is any
qualitative change in motion, as opposed to a detailed quantitative
one, for a small change in energy or in the initial conditions.  A
small change in the amplitude of vibration due to a small change in
energy is like a continuous deformation of the tajectory.  In
topology, the rule of deformation is to bend or stretch in whatever
way we want except that neither distinct points be identified (no
gluing) nor any tearing be done.\footnote{Why these restrictions?  We
  shall see in Sec. \ref{sec:spaces} that ``gluing'' is an equivalent
  relation that changes the space.  Similarly tearing changes the
  space by redefining the neighbourhoods at the point of cut.}  Such
transformations are called continuous transformations.\footnote{See
  Sec. \ref{sec:continuity-function} for a discussion on continuous
  functions}

If the energy is changed continuously from $E_0$ to $E_1$ by defining
a continuous function $E(\tau)$, say $E(\tau)=E_0+ (E_1-E_0) \tau$
with $\tau \in[0,1]$ do the trajectories in phase space get deformed
continuously?\footnote{This is a virtual change and not a real
  time-dependent change of energy. At every $\tau$, the pendulum
  executes the motion for that energy} The answer is not necessarily
yes.  This is where the global properties of the phase space or the
configuration space come into play.  The continuous deformations then
help us both in characterizing the phase space and in classifying the
trajectories.  We show that the three types of motion, O, R and C
belong to three different classes of curves in the appropriate phase
space.

The topological analysis is done by identifying (i) the configuration
space and the phase space, (ii) the possible trajectories on these
spaces, and for Hamiltonian systems, (iii) the constant energy
``surface'' (or manifolds) for possible real motions.  We do these
qualitatively first and then discuss some of the features in more
detail.

\figphspace

\subsubsection{Configuration space and phase space:} 
The first step is to construct the configuration and the phase spaces.
The values taken by the degrees of freedom defines a set.  One
then defines a topology on it by defining the open sets, thereby
generating the topological space to be called the configuration space.
The phase space is obtained by adding the momenta variables to the
configuration space.

For a pendulum, with $ q$ an angle, the configuration space is a
circle $S^1$.  It is also possible to represent the configuration
space as the real line ${\mathbb{R}}$ with an identification of all
points $x=x+2n\pi$, $n\in \mathbb{Z}$ as in Fig. \ref{fig:pend1}.  The
momentum, being real, trivially belongs to ${\mathbb{R}}$, the real
line.
The phase space is therefore $ S^1\times {\mathbb{R}}$.
The direct product phase space
is the surface of a {\it cylinder} (see Fig.  \ref{fig:pend2}), where
the radius of the cylinder is not important.

Note that the phase space can also be viewed as the extended space
${\mathbb{R}}\times {\mathbb{R}}$ with proper interpretation of one
${\mathbb{R}}$.

\subsubsection{ Trajectories:}
Since the generalized coordinates and the conjugate momenta change
continuously with time, the motion of the bob  generates a curve in the phase
space.  This curve is called a trajectory.  The second step is to
construct all the possible trajectories.

As Eq.  (\ref{eq:1}) is time reversible, any piece of trajectory
in the upper half plane of the $(q,p)$ phase space with arrow to the
right (indicating the direction of motion),  has  a mirror image in
the lower half ($p\to -p$) with arrow towards left.  As a corollary,
this time-reversed pair meets to form {\it a closed orbit}, if and
only if there is a point with $p=0$.  In addition, uniqueness of
solution for a given initial condition forbids crossing of distinct
trajectories.

\figcylorb

Let us now combine all the above features.  We find that the possible
trajectories in the extended ${\mathbb{R}}^2$ space are either closed
loops around $ q=2n\pi$ or open curves.  See Fig.  \ref{fig:pend3}.
There is the transitional one that connects the saddle points at
$q=(2n+1)\pi$ at $p=0$.  On the cylindrical phase space, all are
closed orbits; one type (for $E<2$) enclosing the stable fixed point
$(0,0)$, and another type encircling the cylinder (with $E>2$).  The
latter can be grouped into two inequivalent classes, namely the time
reversed partners (see item 6 in Sec. \ref{sec:mechanics}).  It is
obvious that one type cannot be deformed into the other one if we
follow the rules of deformation on the cylinder.  The special one is
$E=2$, a conjoined twin connected at one point. It requires a pinching
(i.e., identification) of two points on the curve for $E<2$.  A
tearing is required as $E$ exceeds 2 by any amount no matter how
small.  Moral: The topology of the phase space naturally separates the
three types of orbits.  The cylindrical topology forbids
transformation of one to the other.

Another way to see the change is to use  energy $E(p,q)$ as a
parameter or replace $p$ by $E$.  As $E$ depends quadratically on $p$,
the cylindrical phase space becomes a U-tube.  To be noted that the
horizontal $p=0$ circle on the cylinder has now become the vertical
circle in the middle of the U as the minimum energy is zero for
$\theta=0$ but 2 for $\theta=\pi$.  The motion is then given by the
intersection of the U-tube with an $E=$const plane. See Fig.
\ref{fig:pend3}.  The three classes of closed loops are now easy to
see.  The corresponding orbits in the configuration space are shown in
Fig. \ref{fig:pend2}.

The peculiarity of the critical case is revealed by the response of
the pendulum to a vanishingly small random perturbation at say the
turning or the top point.  There will be no drastic change for the
$E>2$ or the $E<2$ cases.  But for the figure eight case, when $E=2$,
the motion would consist of any combination of clockwise (C) and
anticlockwise (A) orbits like CCAAAACACCC... . That is to say, all
infinitely long two letter words are possible trajectories, and any
two words differing in at least one letter are distinct. 

\begin{problem}
  Suppose acceleration due to gravity $g=0$.  The phase space is still
  a cylinder but the motion is different.  Discuss how the topological
  arguments change, by focusing on the change in the U-tube for energy
  as $g\to 0$.
\end{problem}

\section{Topological analysis: details}\label{sec:topol-analys-deta}
We now discuss how topology is used in the description.  Let us
remember that a topology on a set $X$ of points require a set of
subsets, $\tau$, to be called open sets, such that (i) $\emptyset{}$
(Null set) and full set $X$ are open, i.e., $\emptyset{},X\in \tau$,
(ii) any finite or infinite union of open sets is also open, 
(iii)  any {\it finite} intersection of open sets, i.e. members of $\tau$,
belongs to $\tau$.  Under these conditions, the set of subsets,
$\tau$, is called a topology on $X$, while $(X,\tau)$ is said to
constitute a topological space.  See Ref. \cite{somnath}.

A useful procedure to define a topology on a set is to embed it in a
known space.  Then use the intersections of the open sets of the known
space with the set in hand to define the open sets in it. For example,
$S^1$ can be drawn in a two dimensional space and the open sets on
this curve can be defined as the intersection of the curve with the
rectangles (or Disks).  The open sets on the circle then form the
basis for the topology on $S^1$.  The topology thus defined is called
the {\it Inherited topology or the subspace topology}.  Since we shall
mostly be working with these inherited topologies, we shall not be
explicit about it anymore, unless something else is meant.  Such
embeddings are useful in most physics problems but there are many
cases for which no embeddings are possible.

Right now we rely on our intuition of spaces.

\subsection{Configuration Space}
\label{sec:spaces}
Our familiarity with the real $d$-dimensional Euclidean space $R^d$
allows us the luxury of thinking of other spaces in terms of $R^d$.
With that, it might be possible to make topological identifications of
spaces.  There should be a  statutory warning that proving the
topological equivalence of spaces in general could be a notoriously
difficult problem.

\subsubsection{$S^1$ as the configuration space}
\label{sec:s1-as-configuration}
Our intuition of angle leads us naturally to $S^1$.  If we think of
the set of values $q\in [0,2\pi]$, we get $S^1$ only if we identify
$0$ and $2\pi$.  This can be seen by gluing the two ends of a piece of
a string (i.e. implement the ``{\it periodic boundary condition}'').
A simple but concrete way of seeing this is to note that a continuous
map takes $q\to z=e^{i q}$ with $z$ defining the unit circle in the
complex plane.

\subsubsection{${\mathbb{R}}$ as the configuration space: equivalence
  relation, quotient space}
\label{sec:r-as-configuration}
A slightly different identification is required for the extended real
line used in Fig.  \ref{fig:pend1}.  This involves an {\it equivalence
  relation} that any $x\in {\mathbb{R}}$ is equivalent to all points
$x+na$ for $n\in Z$.  It is like a translational symmetry in one
dimension.  Any point on ${\mathbb{R}}$ can then be brought into the
interval $[0,a]$ or $[-a/2,a/2]$ by addition or subtraction of
suitable multiples of $a$.  E.g., $-a/2=a/2 -a$ sets $-a/2$ as
equivalent to $a/2$.  This finite closed interval with end point
identification is $S^1$.  If the equivalence is denoted by the symbol
$\sim$, i.e. $x\sim x+na$, then under this equivalence condition, the
relevant space is different from $\mathbb{R}$.  It is denoted by
${\mathbb{R}}/\!\!\sim$, and is called the {\it quotient space } for
the equivalence relation $\sim$.  The space obtained via an
equivalence as ${\mathbb{R}}/\!\!\sim$ is topologically equivalent to
$S^1$.  A general feature we see here is the possibility of
construction of new spaces from a given space by defining an
equivalence relation on it.

\paragraph{{\bf{}}More examples}
For the closed interval $I=[0,1]$, we may define the periodic boundary
condition as an end point equivalence relation $0\sim 1$.  Then
$I/\!\!\sim$ is homeomorphic to $S^1$, where {\it homeomorphism} is
synonymous to ``topologically identical''.  To be more systematic, one
defines (i) a direct map (i.e. a function) $f: I\to S^1$ as
$f(x)=e^{i2\pi x}$, and  successive maps (ii) a map $f_1: I\to
I/\!\sim$, and (iii) an inverse map $f_2: I/\!\sim\,\,\to S^1$, so that
$f_2(f_1(x))=f(x)$ for all $x\in I$.

Note that, we {\it do not} get $S^1$ from $[0,1]$ if the end points
are {\it not identified}.  That they are different can be seen by
removing one point from each one of the two sets.  In the closed
interval case we get two disconnected pieces whereas for $S^1$ we get
an open interval.  If we take an open interval\footnote{The standard
  convention is to use parenthesis $(,)$ to denote openness.  Here the
  boundary points $a,b$ are not in the set $(a,b)$.} like $q\in
(0,1)$, then it is actually equivalent to the whole real line as one
may verify by the map $x\to X=\tan[\pi(x- \frac{1}{2})]$ with $X\in
(-\infty,+\infty)$.

\begin{problem}
  The change in the topological space by an equivalence relation has
  important consequences in physics too.  Take the case of $[0,1]$ and
  $S^1$ under the equivalence condition.  For single particle quantum
  mechanics, the first case corresponds to the boundary condition
  where the wave function $\psi(x)=0$ at $x=0,1$ while the
  second one to periodic boundary condition with 
  $\psi(0)=\psi(1)$.

Take the conventional momentum operator $-i\hbar d/dx$ with the eigen
value equation $-i\hbar \frac{d\psi}{dx}=p\psi$.  Show that  there  is no  valid
solution (i.e., $p$ is not real) for the $[0,1]$  case while $p$ is real for $S^1$.
\end{problem}

\subsubsection{Pendulum vs harmonic oscillator: }
\label{sec:pendulum-vs-armonic}
The importance of the topology of the configuration space can be
understood by comparing the $S^1$ case with the space for the
linearized simple pendulum.  For the latter, the configuration space
is just a small part of the circle (small angles), which can be
extended to the whole real line as for a linear harmonic oscillator.
For this space ${\mathbb{R}}$, there is only one stable fixed point at
$x=0$, and the phase space has only one kind of orbit, namely the
closed orbit of libration type.  All the richness of the full pendulum
at various energies come from the nontrivial topology of the
configuration space.

\subsection{Phase space}
\label{sec:phase-space}

Now that we know the configuration space, we may go over to the phase
space.  The momentum part is easy - it is a real number for the
pendulum, $p\in (-\infty,+\infty)$, i.e. $p\in {\mathbb{R}}$.  For $n$
degrees of freedom there are $n$ momenta, and so the space for the
momenta is ${\mathbb{R}}\times {\mathbb{R}}...\times
{\mathbb{R}}={\mathbb{R}}^n$.  Since the momentum and the position are
independent variables, we have a product space of
${\mathbb{R}}^n\times C$ where $C$ is the configuration space, as seen
for the planar pendulum.

The motion of the pendulum is restricted to the constant energy
subspace of the phase space as shown in Fig. \ref{fig:pend3}.  Beyond
visualization, the differences in the nature of the spaces show up
through their topological invariants, e.g., by the fundamental group.

\subsubsection{Topological invariants --- homotopy groups:}
\label{sec:topol-invar-homot}

One way of exploring a topological space $X$ is by mapping known
spaces like circles, spheres, etc in $X$.  The  case with circles
tells us how many classes of nonequivalent closed loops that start and
end at a point $x_0$ can exist in $X$. Any point in $X$ can be chosen
as the base point $x_0$.  Two loops that can be deformed into one
another are called homotopic\cite{samik}.  All such homotopic loops
can be clubbed together as a single class, with any one of them as a
representative one.  One may also define a product of two loops $C_1$
and $C_2$ by going along $C_1$ and then from $x_0$ along $C_2$.  There
will be classes of loops that are homotopic to $C$, and therefore the
multiplication is for the classes.  Representing a class by $[\gamma]$
for all loops homotopic to $\gamma$, the multiplication rule can be
written as $[C]=[C_2]\;[C_1]$. An inverse of a loop $C$ can be defined
as the loop traversed in the opposite direction, so that $C\; C^{-1}$
is homotopic to a point (i.e. a trivial loop).  More formally, all
these imply that the closed loops rooted at $x_0$ form a group under
the opration of loop multiplications.  This group is called the {\it
  fundamental group} of $X$, $\pi_1(X,x_0)$.  For a connected space,
i.e. if any two points can be connected by a path in $X$, the special
point, the base point can be chosen arbitrarily. We shall therefore
drop $x_0$ from the notation.

The fundamental group of the space for $E<2$ is
$\pi_1(``E<2")=\mathbb{Z}$ where the base point has been dropped.
This part of the space is like the outer surface of a bowl.  In
contrast, the space for $E>2$ is disjoint -- two tubes, and the loops will depend on
whether the base point, $x_0$, is in one or the other circle.  Although
$\pi_1(``E>2",x_0)=\mathbb{Z}$, but $x_0$ in one circle cannot be
connected by a path to the point on the other.  Such a disjoint set is
characterized by the zeroth homotopy group
$\pi_0(``E>2")=\{-1,1\}\equiv {\mathbb{Z}_2}$ with two elements
signifying two components.  The critical surface is again two circles
but with one common point forming figure 8.  Such a union of spaces
with one common point is called a wedge sum, indicated by a $\vee$.
This fundamental group is now a nonabelian group
$\pi_1(``E=2")=\mathbb{Z}\star\mathbb{Z}$ a free group of two
elements.

The difference in the fundamental group tells us that the spaces are
not identical, i.e., not  homeomorphic.

\begin{problem}
  Show that a square with periodic boundary conditions is equivalent
  to $S^1\times S^1$.  Take a piece of paper and glue the sides
  parallelly.  This is a torus which is associated with two holes.
  Define the appropriate equivalence relation ($\sim$), and convince
  yourself that the compact notation is
  $[0,1]\times[0,1]/\!\sim\,=S^1\times S^1$.  This ``='' means
  ``homeomorphic'' or loosely speaking ``topologically equivalent''.

We have already seen two spaces formed by two circles in the case of a
pendulum.  Compared to  the  disconnected space with
$\pi_0=\{-1,1\}$, a torus has  $\pi_0=0$, just as figure 8
which we obtained from two circles with one point equivalence.   Both
(torus and figure 8) are connected spaces.   That a
torus is not topologically figure 8 is established by $\pi_1(S^1\times
S^1)=\mathbb{Z}\times\mathbb{Z}$ while $\pi_1({\rm figure\
    8})=\mathbb{Z}\star\mathbb{Z}$.
\end{problem}

\begin{problem}
  Discuss the  connection between the fundamental
  groups of the constant energy spaces mentioned above and the real
  trajectories of the pendulum.
\end{problem}
\begin{problem}
The phase space for a  simple harmonic oscillator is $\mathbb{R}^2$
with  $\pi_1(\mathbb{R}^2)=0$.  Discuss the motion with respect to the
corresponding $E$-$x$ surface.
\end{problem}

\begin{problem}
  The energy of a free one-dimensional quantum particle is given by
  $E_k=a k^2$ with the wavevector $k\in {\mathbb{R}}$.  The state with
  the wavefunctin described by $k$ may called left-moving or right
  moving for $k<0$ or $k>0$.  For a particle on a lattice (lattice
  spacing=1), the translational symmetry makes two $k$ values
  equivalent if they differ by a reciprocal lattice vector.  In other
  words the $k$-space is like Fig. \ref{fig:pend1}c.  The equivalence
  relation makes the relevant space $S^1$ as in Fig.
  \ref{fig:pend1}b.  A linear representation of $S^1$ is the interval
  $[-\pi,\pi]$ with the identification of the two end points.  With
  this identification, a right moving particle at $k=\pi$ becomes a
  left moving particle at $k=-\pi$ as defined earlier, but one should
  keep in mind the presence of reciprocal lattice vectors .  Draw
  $E_k$ vs $k$ in this 1st Brillouin zone.
\end{problem}

\begin{problem}
(a) Argue that the configuration space of the pendulum in full
  space (spherical pendulum) is $S^2$ (surface of a three dimensional
  sphere). (b) Discuss the possible types of motions using topological
  arguments.  (c) In spherical coordinates $S^2$ can be described by
  $(\theta,\phi)$ where $\theta\in I=[0,\pi]$, and $\phi\in[0,2\pi]$.
  Why is $S^2$ not a product space $I\times S^1$?
\end{problem}

\begin{problem}
  If all the boundary points of a square are made equivalent, then it
  is topologically equivalent to a surface of a sphere $S^2$.  Take a
  piece of cloth or paper and use a string to bring all the boundary
  points together, as one does to make a bag.  Or take a square and an
  isolated point.  Connect all the points on the boundary to that
  point.
\end{problem}

\begin{problem}
  Bloch's theorem, in solid state physics, is generally proved for a
  lattice with periodic boundary conditions, i.e., on a torus (an
  $n$-torus for an $n$-dimensional crystal.  E.g., a torus is obtained by
  identifying opposite edges of a square.  Note that if all points on the
  boundary of a square are identified (spherical boundary condition) 
  one gets $S^2$.   Is Bloch's theorem  valid for the
  spherical boundary condition?  Are the reciprocal vectors defined for
  the spherical boundary condition? 
\end{problem}

\begin{problem}
  Bulk and edge states: In the tight binding model, a quantum particle
  hops on a square lattice.  Find the energy eigen states under the
  following situations.  Pay attention to bulk and edge states.  (i) A
  particle on $S^1\times S^1$.  In this case there are only bulk
  states.  (ii) With spherical boundary condition, i.e., on $S^2$.
  There are no edges.  But are the bulk states same as in (i)?  (iii)
  Klein Bottle.  No edges but different from (i) and (ii).  (iv)
  Periodic boundary condition in one direction, i.e., on a finite
  cylinder.  There are now two edges.  (v) Antiperiodic boundary
  condition, i.e., a M\"obius strip.  There is now one single edge.
\end{problem}

\begin{problem}
  Argue that the configuration space for the planar motion of a double
  pendulum is $S^1\times S^1$.  If we consider the full three
  dimensional space, then the configuration space is $S^2\times S^2$.
\end{problem}

\begin{problem}
  A classical Hamiltonian system with $n$ degrees of freedom is {\it
    integrable} if there exists $n$ conserved quantities or ``first
  integrals''.  In such a case, the motion is confined on an $n$-torus
  $S^1{\times ...\times}S^1$.  Here the product space indicates that
  the motions can be handled independently.  This is easy to see in
  the action angle variables where the $n$ angles constitute the
  $n$-tori.  Convince yourself about this for the pendulum case and
  for the well-known Kepler problem.  This result is useful in the
  context of the important KAM theorem.
\end{problem}

\begin{problem}
  Kapitza Pendulum: The linearized equation of motion around the
  vertical inverted position of a pendulum under a periodic vertical
  drive is $\ddot{\theta} - (g+a(t)) \theta=0$ where $a(t)=a(t+\tau)$
  is the periodic vertical modulation of the point of suspension.  The
  inverted pendulum is stable if the amplitude of the drive exceeds
  some critical value. This is called a Kapitza pendulum.  Discuss
  the motion of an inverted pendulum under a periodic
  vertical drive vis-a-vis Fig. \ref{fig:pend3}.
\end{problem}

\begin{problem}
  Show that a plane with a hole is equivalent (homeomorphic) to a
  cylinder. With the hole as the origin, use polar coordinates $r,\phi$
  so that $r=0$ is excluded (=hole). Now do a mapping $r\to X=\ln r$
  so $X\in (-\infty,+\infty)$, i.e., $X\in {\mathbb{R}}$ and $\phi$
  defines $S^1$.  Therefore ${\mathbb{R}}^2-\{0\}$ (also written as
  ${\mathbb{R}}^2\string\ \{0\}$) is a cylinder.

Solve the free particle quantum mechanics problem in
${\mathbb{R}}^2-\{0\}$ in  $r,\phi$ coordinates. What are the boundary
conditions?   Do the same on the cylinder by transforming the
Schr\"odinger equation to $X,\phi$ variables.  
The main point of this exercise is to see the importance of one
missing point that changes the topology of the space.
\end{problem}

\begin{problem}
  Show that a sphere with a hole $S^2\string\ ${N} (N=north pole) is
  equivalent to a
  plane.  The formal proof is by stereographic projection.  A sphere with two
  holes (north and south poles) is like a cylinder, which in turn is a
  plane with a hole.  What about a sphere with three missing points?
\end{problem}

\begin{problem}
  What is the advantage of  going from the cylinder to the extended real plane as
  in Fig. \ref{fig:pend3}?  The real line or plane has the special
  feature that any closed loop can be shrunk to a point.  Such a space
  is called simply-connected (as opposed to multiply-connected as in
  the previous problem).  A practical usefulness may be seen by
  considering a damped pendulum described by $ \ddot{\theta}+\gamma
  \dot{\theta} +\sin\theta=0,$ where $\gamma$ is the friction
  coefficient.  Now energy is not conserved, $dE/dt<0$, so that the
  pendulum ultimately for $t\to\infty$ comes to rest at the stable
  fixed point $\theta=0$.  Draw the possible trajectories (phase
  portrait) of this damped pendulum for different values of $\gamma$
  and starting energy ($E>2,E<2$) both on the cylinder and on the
  extended space.
\end{problem}

\section{Topological spaces}
\label{sec:topology-space}
 
Is the combination of two real variables $q,p$ equivalent to a two
dimensional Euclidean plane?  The question arises because even if we
take $q$, and $p$ as the two directions of the xy plane, still we may
not be in a position to define a distance between two points
$(q_1,p_1)$ and $(q_2,p_2)$.  The second point is that for a physical
system described by two variables, the state space may locally be like
a plane (two dimensional) but different global connectivities may
imply important qualitative differences.  E.g., for a torus and a
sphere, a small neighbourhood of a point may be described by the
tangent plane at that point and would look simmilar, but globally they
are different.  Let us concentrate on the first issue now.

The absence of a metric (or distance) is a generic problem we face
whenever we want to draw a graph of two different parameters.  Take,
for example, a plot of pressure $P$ and volume $V$ for a verification
of Boyle's law.  The plot reassuringly gives us a branch of a
hyperbola, which is defined as the locus of a point such that the
difference in the distance from two fixed points remain constant.  But
it would be ridiculous to define an Euclidean distance between
$(P_1,V_1)$ and $(P_2,V_2)$.  Still, we know, graph plotting does work
marvelously.

The identification is done in steps through topology.  First an
appropriate topological space is defined which can be identified with
the similar topological space in ${\mathbb{R}}^2$.  Then,  use the
equivalence of this topological space and a metric or distance based
${\mathbb{R}}^2$.

Let's start with the real line.  To define a topology we need a list
or a definition of open sets.  Let's define all sets of the type
$(a,b), (b>a)$ and their unions as open sets.  The null set
$\emptyset{}$ and the full set are also members of the set of open
sets.  That these subsets form a topology on ${\mathbb{R}}$ is easy to check.
The set of subsets with the union and intersection rules then defines
a topological space.  For the real line we used only the ``greater
than'' or ``less than'' relation, without defining any distance or
metric.\\[1pt]

\figbasis

This topology can be extended to ${\mathbb{R}}^2={\mathbb{R}}\times
{\mathbb{R}}$ by defining the sets of open rectangles $(a_1,a_2)\times
(b_1,b_2)$. See Fig.  \ref{fig:r2}.    By doing this we defined a
topology in $\mathbb{R}^2$ without using any distance. 
The next step is to define a metric,
the usual Euclidean distance in ${\mathbb{R}}^2$  with which open disks
$D=(x,y|(x-a)^2+(y-b)^2<\epsilon)$ can be defined around a point
$(a,b)$.  It is known that the topology defined by the open rectangles
and their unions is the same as the one defined by the disks.\\[4pt]

By this procedure, with the help of boxes, the $(q,p)$ phase-space can
be taken as a topological space equivalent to ${\mathbb{R}}^2$.  This
equivalence allows one to see all the geometric features of
${\mathbb{R}}^2$ in the graphs we plot or in the phase space, without
explicitly defining the distance.

An important feature of the topology of the phase space is that it is
{\it connected and simply connected}, i.e. one may go from any point
to any other point, and any two paths connecting two points can be
deformed into each other.\footnote{A connected space has zeroth
  homotopy group $\pi_0=0$. A simply-connected space means $\pi_1=0$.}
A connected phase space is nice because that is a sufficient condition
for the applicability of equlibrium statistical mechanics (generally
called {\it ergodicity} - that one can go from any state to any
other).  However, a phase space may as well be in disconnected pieces
in the sense that two parts may be separated by infinite energy
barriers.  Such spaces might be relevant for phase transitions where
the phase space may get fragmented into pieces (``broken ergodicity''
or ordered systems).

\begin{problem}
  In Fig. \ref{fig:r2}, an infinite number of open boxes are used as
  ``basis'' sets to define the topology of ${\mathbb{R}}^2$.  As a
  vector space, we need only two unit vectors ${\bf i,j}$ where the
  number $2$ of ${\mathbb{R}}^2$ determines the number of basis
  vectors.  Where is this ``2'' when defined as the topological space?
  Argue that this dimensionality comes from the number of spaces
  required to construct the boxes.
\end{problem}

\begin{problem}
We defined the  topology for $S^1$ by embedding it in
$\mathbb{R}^2$ (subspace topology).  Is it possible to 
define a topology on $S^1$ without any embedding?
\end{problem}

\figinfb

\begin{problem}
 Consider the one-dimensional motion of a particle in a double
  well $V(x)=\frac{1}{2}K (x^2-a^2)^2$.  See Fig.\ref{fig:infb}(a).
  Discuss the nature of the configuration space and of the phase space.
  Locate the fixed points and draw the phase space portrait.
\end{problem}

\begin{problem}
  Suppose the barrier of the double well potential is infinitely high
  (Fig. \ref{fig:infb}b.)  Argue that the configuration space consists
  of disconnected pieces.  Draw the possible phase portraits.
\end{problem}

\section{More examples of topological spaces}
\label{sec:examples-more-spaces}

Let us now consider a few other well-known spaces used in condensed
matter physics.  Once the spaces are identified, their topological
classifications in terms of fundamental groups and higher homotopy
groups help us in identifying the defects that may occur, the type of
particles that may be seen and so on.  Here we just construct the
spaces in a few cases.

A class of condensed matter systems involve ordered states. like
crystals, magnets, liquid crystals, etc.  These states or phases have
a symmetry described by a group $H$ which is a subgroup of the
expected full symmetry $G$.  For example, the Hamiltonian of $N$
interacting particles
 is expected to be invariant under full translational and rotational
 symmetry, $G$,  but a crystal, described by the same Hamiltonian has
 only a discrete set of space group symmetries. Such phenomena of
 ordering is known as symmetry breaking.  The ordered state is then
 described by a parameter that reflects this subgroup structure of the
 state.    The allowed values of the order parameter  constitutes
 the topological space for the ordered state and this space is called
 the order parameter space.
Of all the ordered states, ferromagnets and liquid crystals are easy
to describe.  We discuss these spaces below.

\subsection{Magnets}
\label{sec:magnets}
A ferromagnet is described by a magnetization vector ${\bf M}$.  In
the paramagnetic phase ${\bf M}=0$ but a ferromagnet by definition has
${\bf M}\neq 0$.  For simplicity (e.g. at a particular temperature or
zero temperature) we take $M=$const , only its direction may change.

Magnets can be of different types depending on the nature of the
vector.  If ${\bf M}$ takes only two directions up or down, then it is
to be called an Ising magnet.  If ${\bf M}$ is a two dimensional
vector, it is an xy magnet, and for a three dimensional vector it is an
Heisenberg magnet.  In the ferromagnetic phase the origin  ($M=0$) is not allowed
and so any vector space  description will be of limited use.  What is then
required is a topological description of the allowed values of the
magnetization.   Since ferromagnetism is a form of ordering of the
microscopic magnetic vectors, we  call the space an  {\it order
parameter space} ${\cal O}$.

It is now straightforward to see that the order parameter spaces
${\cal O}$  are of the following kinds:\\
(i) Ising: ${\cal O}=Z_2$, (0,1) i.e. up or down\\
(ii) xy: ${\cal O}=S^1$ (circle), i.e., the angle of orientation,  $\theta$. \\
(iii) Heisenberg: ${\cal O}=S^2$ (surface of a 3-dimensional sphere),
i.e., angle of orientation, i.e., $\theta,\phi$.\\
(iv) n-vector model: there are situations where the space could be
$S^n,n>2$. 

If we take a macroscopic $d$-dimensional magnet, then at each point of
the sample (${\bf x}\in {\mathbb{R}}^d$) we define a magnetization
vector 
${\bf M(x)}$ or a mapping ${\bf M}: {\mathbb{R}}^d\to {\cal O}$.
That such a mapping can be nontrivial has important implications.
Instead of a fullfledged analysis of the mapping, it helps to see how
loops and spheres in real space map to the orderparameter space.
E.g., when we move along a closed loop in real space, the order
parameter  describes a closed loop in ${\cal
  O}$.  The nature of these closed loops in ${\cal O}$ is given by the
fundamental group $\pi_1({\cal O})$.  A nontrivial $\pi_1$ indicates
there are loops that cannot be shrunk to a point.  This, in turn,
means that if a loop in real space is shrunk, there will be problems
with continuous deformation of the spins; there has to be a singularity
where the orientation of the spin cannot be defined.  These are called
topological defects.  In $d=2$, loops will enclose point defects while
in $d=3$, loops will enclose line defects, with the elements of
$\pi_1({\cal O})$ as the ``charges'' of these defects.

We just quote here the results\cite{samik} that
$\pi_1(S^1)=\mathbb{Z}$, and $\pi_1(S^n)=0, n>1$.  These mean that
only for the xy-magnet there will be point defects in two dimensions
and line defects in three dimensions. In particular, Heisenberg
magnets will have no point (line) defects in two (three) dimensions.
Any Heisenberg spin configuration in real space can be changed to any
other by local rearrangements of spins.  In contrast, for a
2-dimensional xy magnet, a configuration with a point defect of say
charge=1 {\it cannot be converted by local rearrangements of the
  spins} to a defectless configuration.  The question of continuity of
a mapping (i.e. a function) using topology is discussed separately in
Sec.  \ref{sec:continuity-function}.

\begin{problem}
  There seems to be an obsession for $S^n$, but that's not for no
  reason.  Prove that $S^n$ is the only compact
  simply-connected\footnote{A simply-connected space is one where any
    loop can be contracted to a point. This means its fundamental
    group is trivial, $\pi_1=0$.}  ``surface'' in $n$-dimensions
  ($n\geq 2$).\footnote{Any compact, simply connected $n$-dimensional
    ``surface'' is equivalent to $S^n$.  Remember that $S^n$ is the
    surface of a sphere in $(n+1)$-dimensional space, $\sum_{i=1}^{n+1}
    x_i^2=1$. } (Poincar\'e's conjecture.)
\end{problem}

\begin{problem}
  Berezinskii-Kosterlitz-Thouless transition: it is known that the
  cost to create a unit charge defect in the 2-d xy model is
  $\varepsilon \ln L$ for, say, a square lattice of size $L\times L$.
  Since the defect can be anywhere on the lattice, argue that the free
  energy of a single defect at temperature $T$ is $f(T)=\varepsilon
  \ln L - c T \ln L$, where $c$ is some constant.  Take the defect
  free state as of zero free energy. Show that free defects may form
  spontaneously if $T>T_{\text BKT}= \varepsilon/c$.  This phase
  transition is called the Berezinskii-Kosterlitz-Thouless (BKT
  transition).
\end{problem}
\subsection{Liquid crystals}
\label{sec:liquid-crystals}

\subsubsection{Nematics: $\mathbb{R}P^2$}
\label{sec:nematics:-mathbbrp2}

Lest we created the impression that the world is just $S^n$'s, we look
at a different ordered system, namely liquid crystals.  A nematic
liquid crystal consists of rod like molecules where the centres of the
rods are randomly distributed as in a liquid but the rods have a
preferred orientation ${\bf N}$.  This looks like a magnet but it
isn't so because a rod does not have a direction, i.e. it is like a
headless arrow.  A flipping of a rod won't change anything in contrast
to ${\bf M}\to -{\bf M}$.  As a direction in 3-dimensions, the order
parameter space ${\cal O}_{\rm nematic}$ should have been $S^2$ but
not exactly.  Two points on a sphere which are diametrically opposite
represent the same state, and therefore there is an equivalence
relation on the sphere that {\it antipodal points are equivalent}, ${\bf
  N}\to-{\bf N}$.  This is not just the hemisphere but a hemisphere
with the diametrically points identified on the equator.  This is
called the real projective plane $S^2/Z_2={\mathbb{R}}P^2$.  In
general, $S^n/Z_2={\mathbb{R}}P^n$.

As an ordered system, we would like to know how the headless arrows
can be  arranged in space.  This requires the behaviour of the
map  ${\bf N}: {\mathbb{R}}^d\to {\mathbb{R}}P^2$.

\subsubsection{Biaxial nematics}
Instead of rod like molecules, one may consider rectangular
parallelepiped with $2$-fold rotational symmetry corresponding to the
$2\pi$ rotations around the three principal axis.  Such a liquid
crystal is called a biaxial nematics.  The order parameter space is
the sphere $S^2$ with the equivalence relation, $\sim$, under the
three rotations.  This ''$\sim$'' is not just the identification of
the antipodal points but, in addition, the equivalence of four sets of
points (corners of the box) on the surface of the sphere.  The generic
notation ${\cal O}= S^2/\!\!\sim$ is too cryptic to have any use.
This is where the symmetry operations as a group is useful.

The sphere is actually a representation of the rotational symmetry.
If ${{n}}_1, {{n}}_2$ are any two allowed values of the order
parameter, they are related by the three dimensional rotation group
$G=SO(3)$. By keeping any one value fixed, say ${{n}}_1$, all others
can be generated by the application of the group elements of $G$.
However, the special symmetry of the biaxial nematics, a subgroup of
four elements, $H=D_2$, keeps the order parameter invariant, i.e., if
$h\in H$, then $n_1=h n_1$.  Then, there is some $g\in G$, for which
$n_2=g n_1=g h n_1$, so that $n_2$ is generated by all group elements
of the type $gh$ with $h\in H$ and $g\in G$ but not in $H$.  What we
get is the coset of $H$ in $G$, $G/H$.  So, instead of the generic
representation as $S^2/\!\sim$, we may use groups to represent the
order parameter space as a coset space, ${\cal O}= SO(3)/D_2$.  The
similarity of notations (quotient space in topology and coset space in
group theory) is not accidental but is because of the similarity of
the underlying concepts.  It is now straightforward to generalize to
any other point group symmetry.  It will be the corresponding coset
space.

\begin{problem}
Instead of $SO(3)$, one may consider  $SU(2)$.  Under this mapping, show
that $D_2$ goes to a eight member nonabelian group, $Q$, the group of
quaternions.  Therefore, ${\cal O}=SU(2)/Q$.
\end{problem}

\begin{problem}
  Show that the order parameter spaces for magnets can be written in
  terms of groups as the following coset space: 
\begin{enumerate}
\item the xy case: ${\cal O}=SO(2)$,  or ${\cal O}=U(1)$.  Note that
  the coset space is a group in this case.
\item  the Heisenberg case: ${\cal O}=SO(3)/SO(2)$, or ${\cal O}=SU(2)/U(1)$.
\end{enumerate}
\end{problem}

\subsubsection{What is ${\mathbb{R}}P^n$?}
\label{sec:what-rpn}

A real projective space is obtained by identifying the points which
differ by a scale factor.  If any point ${\bf x}\in
{\mathbb{R}}^{n+1}$ is equivalent to $\lambda {\bf x}$ for any real
$\lambda\neq 0$, then under this equivalence relation
$({\mathbb{R}}^{n+1}\string\ \{0\})/\!\sim={\mathbb{R}}P^n$.
Geometrically, all points on a straight line through the origin are
taken as equivalent.  The space then consists of unit vectors ${\bf
  n}$ with ${\bf n}$ equivalent to $-{\bf n}$.

\begin{problem}
What is the configuration space of a rigid diatomic molecule in 3
dimensions?  \\
Ans: ${\mathbb{R}}^3$ for the centre of mass and $S^2$ for orientation of the
molecule.
In case the two atoms are identical then it is ${\mathbb{R}}^3\times {\mathbb{R}}P^2$.
\end{problem}

\figtwoc

\paragraph{${\mathbb{R}}P^1$:}
\label{sec:rp1}

Take a circle and identify the diametrically opposite points.  See
Fig. \ref{fig:twoc}a.  This is
easy to do with a rubber band.  The folding process shows that
$S^1/Z_2={\mathbb{R}}P^1=S^1$.   Another way of looking at $RP^1$ is
shown in Fig. \ref{fig:twoc}b. Take all straight lines through origin
in two dimensions.   Then declare all points on a line, except the origin,  as
equivalent.   We may choose any point (not origin) on a  line as a
representative point.  Draw a circle through the origin with the
center on the y-axis.  Every line meets this circle at a point (again
exclude the origin) which can be taken as a representative point of
the line. There is therefore one-one correspondence between the
points on the circle and the lines through the origin.  The excluded
point on the circle (the origin of ${\mathbb{R}}^2$) can then be
included as the representative point for the x-axis. Hence the
topological equivalence of ${\mathbb{R}}P^1$ and $S^1$.

In contrast, ${\mathbb{R}}P^2$ is not simple.  In the Euclidean case any
two straight lines intersect at one and only one point, unless they
are parallel.  Parallel lines do not intersect.  In the real
projective plane, any two straight lines always intersect either in
the finite plane or at infinity\footnote{In paintings, for  proper
  perspective,  parallel lines are drawn in a way that
  gives the impression of meeting at infinity.}.

\subsection{Crystals}
Take the case of a crystal which has broken translational symmetry.
If we move an infinite crystal by a lattice vector, the new state is
indistinguishable from the old one.  For concreteness let us take the
crystal to be a square lattice of spacing $a$ in the xy plane.
Consider the atoms to be slightly displaced from the chosen square
lattice. Now, if one atom is at ${\bf r}_0$ from one particular
lattice site, it is at ${\bf r}_{mn}= {\bf r}_0+ m a \hat{i} + n a
\hat{j}$ from a site at $(m,n)$.  All of these are equivalent.  The
order parameter space is then the real plane ${\mathbb{R}}^2$ under
the equivalence condition of translation of a square lattice.  There
is now an equivalence relation that any point ${\bf r}$ is equivalent
to a point ${\bf r}+ m a \hat{i} + n a \hat{j}$ for $m,n\in
{\mathbb{Z}}$.  The order parameter space is therefore a torus.  Note
that this is a generalization of the one dimensional case of Fig.
\ref{fig:pend2} to 2-dimensions, except that we are now going from Fig
\ref{fig:pend2}c to Fig.  \ref{fig:pend2}b.

\subsection{A few Spaces in Quantum mechinics}

We consider the forms of a few finite dimensional Hilbert spaces.
 
\subsubsection{ Complex projective plane $\mathbb{C}P^n$}
\label{sec:compl-proj-plane}

In quantum mechanics, the square integrable wave functions form a
Hilbert space.  Any state $|\psi\rangle$ can be written as a linear
combination of a set of orthonormal basis set $\{|j\rangle\}$ as
$|\psi\rangle=\sum_j c_j\, |j\rangle$.  Let's keep the number of basis
vectors finite, $n<\infty$.  The set of $n$ complex numbers $\{c_j\}$
is an equivalent description of the state so that the state space is
an $n$-dimensional complex space ${\mathbb{C}}^n$.  Since only
normalized states matter, $\{c_j\}$ and $\{\lambda c_j\}$ for any
complex number $\lambda$ represent the same state.  Hence there is an
equivalence relation $\{c_j\} \sim \{\lambda c_j\}$ in $\mathbb{C}^n$.
The relevant space for wave functions is then 
$(\mathbb{C}^n{\string\ } \{0\})/\!\!\sim\,=\mathbb{C}P^{n-1}$, 
a complex projective space in
analogy with real projective spaces.

\subsubsection{Two state system}
\label{sec:two-state-system}

A particular case, probably the simplest, is a two level system (a
qubit), like a spin 1/2 state with $|+\rangle$ and $|-\rangle$ states.
The space of states is therefore the one-dimensional complex
projective plane $\mathbb{C}P^1$.  Any normalized state can be written
as
\begin{equation}
  \label{eq:7}
 |\psi\rangle=\cos\frac{\theta}{2}\;\; |+\rangle + e^{i\phi}
\sin\frac{\theta}{2}\;\; |-\rangle, \;\; {\rm with\;\; } 0\leq \theta\leq\pi,
0\leq\phi\leq 2\pi.  
\end{equation}
The two angular parameters $\theta$ and $\phi$ allow us to map
${\mathbb{C}}P^1$ to $S^2$, called the {\it Bloch sphere}, though
there are problems with the representation of $|+\rangle$ and
$|-\rangle$.  To get rid of this problem one actually needs two maps.
From the equivalence relation $(c_1,c_2)\sim (\lambda c_1,\lambda
c_2)$, we may choose $\lambda$ to write $(c_1,c_2)\sim (1,z)$ or
$(c_1,c_2)\sim (1/z,1)$ so that the two original basis vectors
$|\pm\rangle$ come from $z=0$ or $z=\infty$.  The addition of the
point at infinity to the complex plane gives us the {\it Riemann
  sphere}, also called {\it one point compactification of the complex
  plane}.

The sphere allows us to define a metric in terms of $\theta,\phi$,
which then acts as a metric, the Fubini-Study metric, for
$\mathbb{C}P^1$.

\subsubsection{Space of Hamiltonians for a two level system}
\label{sec:space-hamilt-two}

The Hamiltonian for a two state system is a $2\times 2$ Hermitian
matrix.  Therefore we may consider the space of all such Hamiltonians.
Any Hermitian $2\times 2$ matrix can be expressed in terms of the
Pauli matrices\footnote{Pauli matrices are taken in the standard form,
  where $\sigma_y$ is complex, as
  \begin{equation}
    \sigma_x=\left( \begin{array}{cc}
          0 & 1\\
           1         &0
           \end{array}\right), 
 \sigma_y=\left( \begin{array}{cc}
          0 & -i\\
           i         &0
           \end{array}\right),
 \sigma_x=\left( \begin{array}{cc}
          1& 0\\
           0        &-1
           \end{array}\right).
  \end{equation}
}
\begin{equation}
  \label{eq:12}
  H=\left( \begin{array}{ll}
          \epsilon_1 & a-ib\\
           a-ib         &\epsilon_2
           \end{array}\right) 
= \frac{\epsilon_1+\epsilon_2}{2} \mathbb{I} + a\; \sigma_x+ b\; \sigma_y+\frac{\epsilon_1-\epsilon_2}{2}\;\;\sigma_z.
\end{equation}
In general, the space of the $2\times 2$ Hermitian Hamiltonians is a
real four dimensional space with $({\mathbb{I}}, \sigma_x, \sigma_y,
  \sigma_z)$ as the basis vectors..  It has to be a real space because
hermiticity requires all the four numbers, $\epsilon_1, \epsilon_2,
  a, b$ to be real.

In some situations a further reduction in the dimensionality of the
space is possible.  
By a shift of origin, we may set $\epsilon_1=-\epsilon_2=\epsilon$ to
make $H$ traceless.   In this situation,
\begin{equation}
  \label{eq:13}
  H={\bf d}\cdot {\bf\sigma}=|d| {\bf n}\cdot {\bf\sigma},
\end{equation}
with ${\bf n}={\bf d}/|d|$, a unit vector.  The set of all such
traceless Hamiltonians can   be described by the vector ${\bf n}$, {\it
  provided} $|d|\neq 0$.  Therefore, the space of the Hamiltonians of
any two level system is $S^2$.  The center of the sphere corresponds
to the degenerate case, $|d|=0$, when the two energy eigenvalues are
same.

A practical example is a spin-$1/2$ particle in a magnetic field with
$H=-{\bf h}\cdot {\bf {\sigma}}$,  where $h$ may depend on some
external parameters including time.  Another example is a two band
system.  For a one dimensional lattice, consider two bands
$\epsilon_1(k),\epsilon_2(k)$ with some symmetry such that
$\epsilon_1+\epsilon_2=$const for all $k$.  Choosing the constant to
be zero, we now have  Hamiltonian of the type Eq. \ref{eq:13}
 with ${\bf d}(k)$ a function of the quasimomentum $k$, where $k$ is in
 the first Brillouin zone, $-\pi\leq k\leq \pi$.  We therefore have a
 map $S^1\to S^2$.  In two dimensions, the Brillouin zone is a torus
 and therefore we need to study the map ${\mathbb{T}}^2\to S^2$.
Some aspects of these maps are considered in Sec. \ref{sec:quantum-two-level}.

\begin{problem}
  Construct the topological space for the Hamiltonian of a three level
  system.  Explain why it is reasonable to expect $SU(3)$ and not a
  spin $s=1$ state.  Generalize it to $m$-level system for any $m$.
\end{problem}

\begin{problem}
  The Bloch sphere describes the pure states.  The density matrix of a
  state $|\psi\rangle$ is $\rho=|\psi\rangle\;\langle\psi|$, with
  $\rho^2=\rho, \ {\rm Tr}\ \rho=1$.  These two conditions on $\rho$
  can be taken as the definition of a pure state without any reference
  to wave functions.  In this scheme, mixed states are those for which
  ${\rm Tr}\ \rho=1$, but $\rho^2\neq \rho$.  This means $P={\rm Tr}\
  \rho^2 <1$.  $P$ is called the purity of the state.  For the two
  state system, mixed states are given by $2\times 2$ Hermitian,
  positive semidefinite\footnote{Positive semidefinite means all the
    eigenvalues, $\lambda_i$'s satisfy $\lambda_i\geq 0$.  For a
    density matrix we need $0\leq\lambda_i\leq 1, \sum_i\lambda_i=1$.}
  matrices with trace 1.  Show that these mixed states are points
  inside the Bloch sphere.  The relevant space is now a 3-ball (a
  solid sphere).
\end{problem}

\section{Disconnected space: Domain walls}
\label{sec:disc-conf-space}

Of all the order parameter space for  a magnet defined in Sec
\ref{sec:magnets}, the Ising class is special because here the space is
disconnected.  
The same result is obtained by using  the $\phi^4$ theory with an energy
functional 
\begin{equation}
  \label{eq:4}
  E[\phi(x)]=\int_{-\infty}^{\infty} dx\  \left[ \frac{1}{2}
    \left(\frac{d\phi}{dx}\right)^2+ \frac{1}{2} K \left(\phi(x)^2-\phi_0^2\right)^2\right],
\end{equation}
so that the minimum energy states correspond to $\phi(x)=\pm \phi_0$.
For finite energy states,  we require $\phi(x)$ to be
nonconstant but going to  $\pm\phi_0$ as $x\to\pm\infty$.  The
requirement at infinity gives us four possibilities, shown in a
tabular form below.

\begin{table}[h]
  \label{tab:tab:1}
\begin{minipage}{0.25\textwidth}
\begin{tabular}{|l|c|c|}
\hline
\quad&$\phi|_{x\to -\infty}$& $\phi|_{x\to\infty}$\\
\hline
(a) &$\phi_0$& $\phi_0$\\
\hline
(b) & $-\phi_0$&$-\phi_0$\\
\hline
(c) &$-\phi_0$&$\phi_0$\\
\hline
(d) &$\phi_0$&$-\phi_0$\\
\hline
\end{tabular}
\end{minipage}
\begin{minipage}{.35\textwidth}
  \caption{Boundary conditions at $\pm\infty$.}
\end{minipage}
\end{table}

These four cases are distinct because there is no continuous
transformation that would change one to the other.

For cases (a) and (b), local changes (like spin flipping) can reduce
the energy to zero and these represent small deviations from the
fully ordered uniform state of $\phi_0$ or $-\phi_0$. These two states are
related by symmetry but they are distinct.

For cases (c) and (d), no continuous local transformation can change
the boundary conditions to the uniform state.  Therefore, they
represent different types of states.  These finite energy states are
called topological excitations because their stability is protected by
topology.  This is a domain wall or interface separating the two
possible macroscopic state $\pm\phi_0$.  These topological excitations
are  called  kink for (c) and anti-kink for (d).

More generally for any discrete or disconnected configuration space,
i.e., if its $\pi_0$ (zeroth hommotopy) is nontrivial, there will be domain walls.
A better description of a disconnected space is via the zeroth
homology, $H_0$, for which we refer to Ref. \cite{dheeraj}.

\begin{figure}[htbp] 
  \begin{center}
\includegraphics[scale=.51,clip]{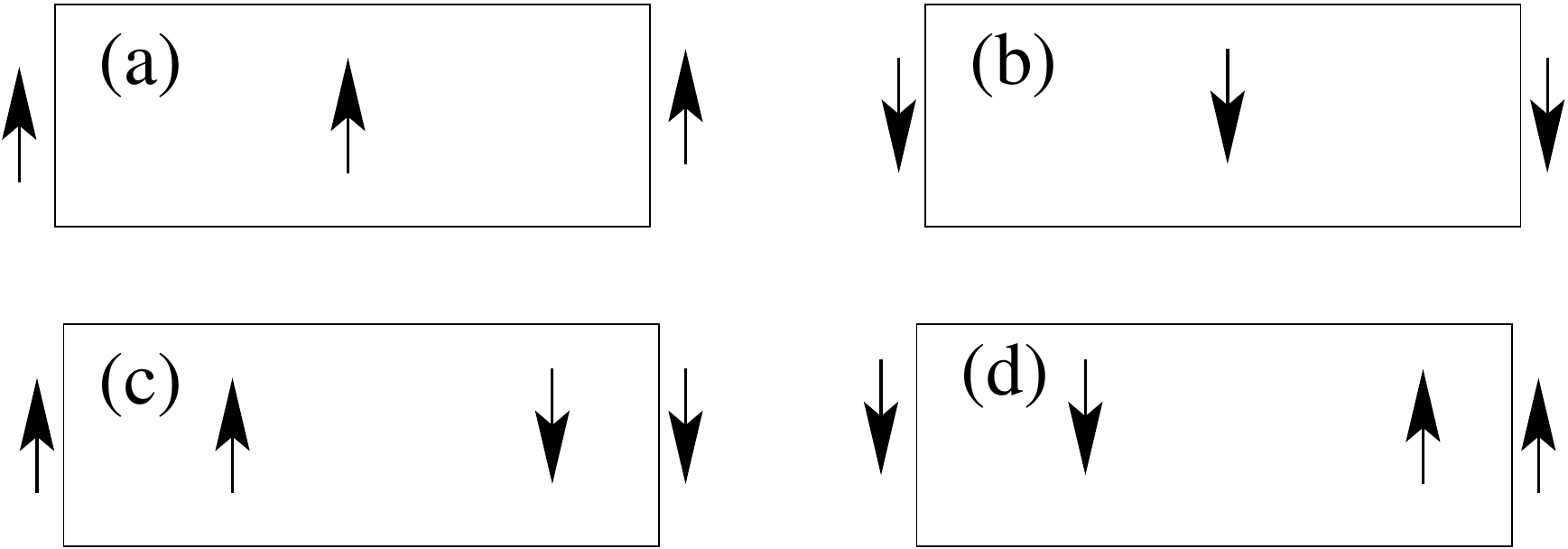}    
  \end{center}
\caption{Four possible boundary conditions for an Ising magnet
with spins=$\pm 1$. The up (down) arrow indicates spin $+1 (-1)$.
An interface exists for (c) and (d).  
}
\label{fig:ising}
\end{figure}

One may see this boundary condition induced domain walls in 
 the ordered state of a two dimensional Ising model on a
square lattice.  If we take a long strip with four different boundary
conditions as in Fig. \ref{fig:ising} along one direction, we force
domain walls in cases with opposite boundary conditions as in
Fig. \ref{fig:ising}c,d.  The energy of the interface is
obtained by subtracting the free energy of (a) or (b) from (c) or
(d). This is ensured in Eq. (\ref{eq:4}) by taking the energy at
infinity to be zero.

\begin{problem}
Use the energy functional of Eq. (\ref{eq:4}) to determine the domain
wall energy.
\end{problem}

\begin{problem} 
  The previous discussion allows for only one type of domain wall to
  exist in a configuration.  A generalization would be to consider a
  case of several disconnected pieces of the configuration space, as
  in the Potts model.  In this model, each ``spin'' can take $q$
  discrete values.  A lattice Hamiltonian with nearest neighbour
  interaction can be of the form $H=- J \sum_{nn} \delta(s_i,s_j),
  J>0$.  Show that the ground state is $q$-fold degenerate.  Discuss
  the nature of domain walls or kinks/antikinks in the Potts model.
\end{problem}

\begin{problem}
The boundary conditions in Fig. \ref{fig:ising} can be classified as
periodic (a,b) and antiperiodic (c,d).  If we join the two vertical
edges (equivalence relation) in a way that matches the arrows, show
that we get a cylinder for (a) and (b) while a Mobius strip for (c)
and (d).  See Appwndix A for a problem on flux thorugh such surfaces.  
\end{problem}

\section{Continuous  functions}
\label{sec:continuity-function}

So far our focus has been on the topological spaces defined for
various sets.  In the process functions are also defined as maps
between two given spaces.  It is necessary to define a continuous
function in topology without invoking the $\epsilon,\delta$
definitions of calculus.

The topological definition of a continuous function is in terms of its
inverse function.  A function $f:A\to B$ is continuous if $f^{-1}$
maps open sets of $B$ to open sets of $A$.  The definition in calculus
is that given any $\epsilon$ no matter how small, if we can find a
$\delta(\epsilon)$, which depends on $\epsilon$, such that
$|f(x+\delta)-f(x-\delta)|< \epsilon$, then $f(x)$ is continuous at
$x$.  In this $\epsilon$-$\delta$ definition, continuity is linked to
closeness as measured by a distance-like quantity.  The topological
definition replaces the neighbourhoods by the open sets, the
constituent blocks of the space, but, in addition, it involves the
inverse function. That should not be a surprise if we recognize that,
by specifying $\epsilon$ for $f$ and then finding $\delta$ for $x$ is
like generating the inverse function.

\newcommand{\figdisc}{%
\begin{figure}[htbp]
\begin{center}
\includegraphics[scale=0.5,clip]{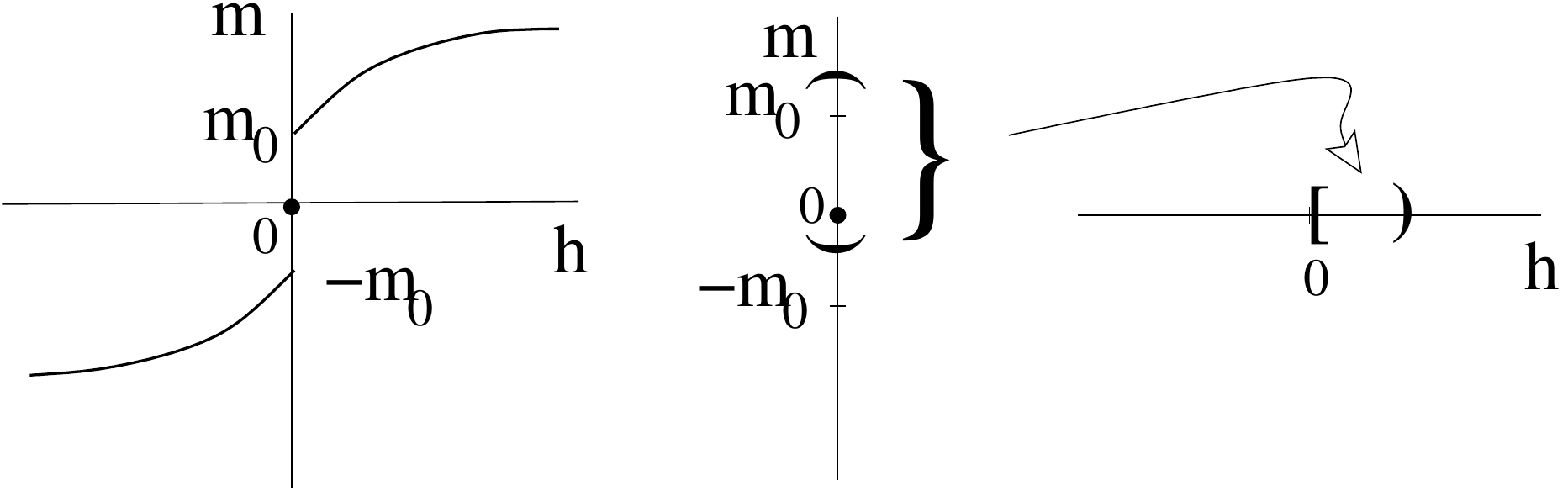}
\end{center}
\caption{A discontinuous function.  Magnetization vs magnetic field
  for a ferromagnet.  At $h=0, m=0$ and $\lim_{h\to0\pm}m(h)=\pm m_0$.
  The example on the right shows an open interval $(-\epsilon,m_1),
  \epsilon>0, m_1>m_0$, maps on to a semiclosed interval $[0,h_1)$
  for $h$, with $m(h_1)=m_1$.  The inverse function maps an open
  interval of $m$ to a semi-open interval of $h$.  }
\label{fig:18}
\end{figure}
}%

\figdisc

We illustrate this with the
help of a known physical example, namely the magnetization of a
ferromagnet in a magnetic field, $m(h)$ with  $m,h\in \mathbb{R}$.
See Fig \ref{fig:18}.
Take any open set $(h_i,h_f)$.    The corresponding values of $m$ form an
open set $(m(h_i),m(h_f))$, missing the ``discontinuous'' nature at
the origin.  In contrast,   the inverse image of $(-\epsilon,m_1)$ maps to the
semiopenset $[0,h_1)$.\footnote{In this one-dimensinal example, an interval has two
  boundaries.  If one boundary is a member of the set but not the the
  other one,  then it is called a {\it semi-open }set. }  

The definition of continuity in terms of the inverse image has
implications in physical situations too.  The response function,
susceptibility, defined as $\chi=\frac{\partial m(h)}{\partial h}$
loses its significance at $h=0$.  The relevant quantity in this
situation is the inverse susceptibility $\chi^{-1}$ which may be
defined as $\frac{\partial h(m)}{\partial m}$ in terms of the inverse
function $h(m)$.  In thermodynamics or statistical mechanics, the
inversion is done by changing the ensemble.  In a fixed magnetic field
ensemble, the free energy $F(h)$ is a function of $h$ while in a fixed
magnetization ensemble the free energy ${\cal F}(m)$ is a function of
the magnetization.  These two free energies are related in
thermodynamics by a Legendre transformation.  By generalizing to free
energy functionals, the two response functions are defined as 
\begin{equation}
  \label{eq:6}
  \chi({\bf r},{\bf r}^{\prime})=\frac{\partial m({\bf r})}{\partial
  h({\bf r}^{\prime})}=-\frac{\delta^2 F}{\delta h({\bf r})\;\delta
h({\bf r}')},\quad 
  \chi^{-1}({\bf r},{\bf r}^{\prime})=\frac{\partial h({\bf r})}{\partial
  m({\bf r}^{\prime})}=\frac{\delta^2 {\cal F}}{\delta m({\bf r})\;\delta
m({\bf r}')}, 
\end{equation}
such that $\int \chi^{-1}({\bf r,r'})\ \chi({\bf r',r''}) d{\bf
    r}' = \delta({\bf r}-{\bf r}'')$.   
Such  inverse response functions lead to  vertex functions in field theories.

\begin{problem}
Consider a field theory like the $\phi^4$ theory of Eq. (\ref{eq:4}),
defined in a $d$-dimensional  space.   Show that the two point
vertex function for this field theory corresponds to an inverse
response function.
\end{problem}

\section{Quantum mechanics}
\label{sec:quantum-mechanics}

A few examples of use of topology in elementary quantum mechanics are
now discussed.  These examples are not to be viewed in isolation but
in totality with all other examples discussed in this chapter We shall
mix classical examples here too to show the broadness of the
topological concepts and topological arguments.  Later on these ideas
in the quantum context will be connected to another, completely
classical, arena of biology involving DNA.

\subsection{QM  on multiplyconnected spaces}
\label{sec:qm-mult-spac}

We now consider quantum mechanics on   topologically nontrivial spaces in 
quantum mechanics, for which we need to reexamine two  traditional
statements, namely, 
\begin{enumerate}
\item Wave functions are single valued.
\item  An overall phase factor, $\psi(x)\to e^{i\phi}\psi(x)$ does not matter.
\end{enumerate}
The singlevaluedness criterion gets translated in the path integral
formulation as the sum over all paths in the relevant configurational
space with weights determined by the action along the path.  These are
actually valid for simply-connected spaces, but not necessarily for a
multiply-connected space.  An example of such a case is the problem of
a single particle on a ring which is discussed in some detail,
avoiding a full fledged general analysis.  The result can be extended
to the case of a plane with a hole (See Prob. 2.13) as we shall do
below.

\subsection{Particle on a ring}
\label{sec:particle-ring}

A particle is constrained to move on a ring (a circle) of circumference $L$. We
use the coordinate $x$ to denote the position on the ring. 

For a circle $S^1$ and its universal cover $\mathbb{R}$, refer to Fig.
\ref{fig:pend1}b,c and Fig ~\ref{fig:pend2}b,c.  To maintain
generality, we use  notations $\mathbb{F}$ and
$\widetilde{\mathbb{F}}$ as the topological space in question, and its
universal cover respectively.  These are related by
$\mathbb{F}=\widetilde{\mathbb{F}}/G$, where $G$ is a discrete group
expressing the equivalence relations.  For the ring, $\mathbb{F}=S^1$,
and $\widetilde{\mathbb{F}}={\mathbb{R}}$, and $G=\mathbb{Z}$ (See
Sec. \ref{sec:spaces}).  The requirement that $
\pi_1(\widetilde{\mathbb{F}})=0$, sets $\pi_1(\mathbb{F})=G$.

A trivial loop in $\mathbb{F}$, see Fig. \ref{fig:pend2}, maps to a
loop in $\widetilde{\mathbb{F}}$, whereas all nontrivial loops in
$\mathbb{F}$ become open paths in $\widetilde{\mathbb{F}}$.  A point
$q$ in $\mathbb{F}$ maps to many points (all equivalent) in
$\widetilde{\mathbb{F}}$.  Let us choose one such point
$\widetilde{q}_0$ arbitrarily. A closed loop $C$ in $\mathbb{F}$ from
$q_0$ to $q_0$ maps to a unique path $\widetilde{C}$ in
${\widetilde{\mathbb{F}}}$ from ${\widetilde{q}}_0$ to another
equivalent point ${\widetilde{q}}^{\prime}_0$.  The end point is the
same for all paths homotopic to $C$ (i.e.  deformable to $C$).  By
denoting all such homotopic paths by $[C]$ or $[\widetilde{C}]$ as the
case may be, we write ${\widetilde{q}}^{\prime}_0=[C]\;
{\widetilde{q}}_0$, without using any tilde on $C$.

It is now reasonable to expect
\begin{equation}
  \label{eq:quan1}
\widetilde{\psi}([C]\widetilde{q})=a([C])\;\widetilde{\psi}( \widetilde{q}),  
\end{equation}
with $a$ as a phase factor.  Two loops $C,C'$ in $\mathbb{F}$ based at
$q_0$ can be combined into one\footnote{This rule, in fact, generates
  the fundamental group $\pi_1(\mathbb{F})$.}, $[C'']=[C']\; [C]$.  In
$\widetilde{\mathbb{F}}$ the corresponding paths $\widetilde{C}$
connects $\widetilde{q}_0$ to $[C] \widetilde{q}_0$ while the
subsequent $\widetilde{C}'$ connects $[C]\widetilde{q}_0$ to $[C']
[C]\widetilde{q}_0$ which is also $[C'']\widetilde{q}_0$.  For the
wavefunction, we get the combination rule for the phase factors as
\begin{equation}
  \label{eq:quan2}
  a([C']) a([C])=a([C'] \; [C])=a([C'']),
\end{equation}
i.e., $a$'s follow the group multiplication rules of $\pi_1$.  These $a$'s
therefore constitute  a one-dimensional representation of the fundamental
group.

A simple path in $\mathbb{F}$ connects any point $q$ to $q_0$.  Simple
here means the path shrinks to a point as $q\to q_0$.  Such paths
allow us to map all the points of $\mathbb{F}$ to a domain containing
$\widetilde{q}_0$ in $\widetilde{\mathbb{F}}$.  For the circle case,
this is reminiscent of a unit cell in $\mathbb{R}$.  The equivalence
relation or the nontriviality of $G$ suggests that there are other
equivalent domains, as many as the number of elements of $G$.  As an
example, in Fig. \ref{fig:pend1} , the domains can be chosen as
$(-\pi,\pi], (\pi,3\pi]...$, an infinite of them as $\mathbb{Z}$ is
countably infinite.  The wavefunction is single valued in each of
these domains but those in two different domains differ by a phase
factor.  As any of these domains is isomorphic to ${\mathbb{F}}$, any
one of these wavefunctions can be taken as the wavefunction on
$\mathbb{F}$.  We end up with a multivalued wavefunction on
$\mathbb{F}$ whose branches are the wavefunctions on the ``unit
cells'' of $\widetilde{\mathbb{F}}$.  In short, quantum mechanics on a
multiply connected space requires a multivalued wavefunction, unlike
the simple cases studied in Euclidean spaces.\footnote{An analogy: In
  the complex plane, $f(z)=\sqrt{z}$ is multivalued but it is
  single-valued on the extended Riemann sheets.  Each sheet defines
  one branch of $f(z)$.  Compare this with multivalued $\psi(q)$ on
  ${\mathbb{F}}$ but single-valued $\widetilde{\psi}$ on
  $\widetilde{\mathbb{F}}$.}  This, fortunately, is not the end of the
story.  With the help of examples, we shall see that we may still
choose single valued wavefunction,  at the cost of an extra phase though.
This is Berry's phase which goes beyond topology and appears in many
problems as a geometrical phase  

Let us consider a few special cases.

\subsubsection{  $a=1$: single-valued wavefunction}
\label{sec:a-sing}

The identity representation is the trivial representation of any group.
Let us choose $a=1$ for all elements of $G=\mathbb{Z}$.   The free
particle Hamiltonian 
\begin{equation}
  \label{eq:quan5}
 H=\frac{p^2}{2m}, \quad {\rm with\ } H \psi(x)=E\ \psi(x),\quad {\rm
   and\;\;\; }\psi(0)=\psi(L).
\end{equation}
The periodic boundary condition, which incorporates our requirement of $a=1$,  
  gives the known energy eigenfunctions and eigenvalues as
\begin{equation}
  \label{eq:quan3}
 \psi_k(x)=e^{ikx}.\quad   k=\frac{2\pi n}{L}, n\in \mathbb{Z}, \ {\rm and} \ E_n=\frac{2\pi^2 \hbar^2n^2}{mL^2}. 
\end{equation}
Importantly, the wavefunction is single-valued.

\subsubsection{  $a_n=e^{in\theta}$: multi-valued wavefunction}
\label{sec:a_n-mult-gauge}

Let us now consider the case of multivalued wavefunction.  By using
gauge transformation, the multivaluedness is linked to the behaviour
of a particle when the ring is threaded by a  magnetic flux.  A
connection between the two problems is then obtained via Berry's phase.

\paragraph{A.  Multi-valuedness\\ }
\label{sec:multi-valuedness}

Let us  choose, respecting Eq. (\ref{eq:quan2}),  a unitary representation
\begin{equation}
  \label{eq:quan10}
 a_n=e^{in\theta}, n\in\mathbb{Z}, 
\end{equation}
It is, as per Eq. (\ref{eq:quan1}),  equivalent to a twisted  boundary condition 
\begin{equation}
  \label{eq:quan12a}
\psi(0)=e^{-i\theta}\psi(L),  
\end{equation}
thereby making the wavefunction multivalued.

The energy eigenvalues and eigenfunctions are still given
by  Eq. (\ref{eq:quan3}) but with 
\begin{equation}
  \label{eq:quan7}
  k=\frac{2\pi n+\theta}{L}, \ {\rm and} \ E_n=\frac{2\pi^2 \hbar^2(n+n_0)^2}{mL^2}, \quad n\in \mathbb{Z},
\end{equation}
 where $n_0=\theta/(2\pi)$.

\paragraph{B.  Gauge transformation, magnetic flux\\ }
\label{sec:gauge-transformation}
We may do a gauge transformation for the wavefunction,
$\Psi(x)=e^{i\Theta(x)} \psi(x)=U \psi(x)$, where $U$ is the unitary
transformation operator.   Such a transformation changes the boundary condition
to
\begin{equation}
  \label{eq:quan8}
  \Psi(L)=e^{i\Theta(L)}\psi(L)=e^{i\Theta(L)} e^{i\theta} \psi(0)=e^{i\Theta(L)} e^{-i\Theta(0)} e^{i\theta} \Psi(0).
\end{equation}
The choice 
\begin{equation}
  \label{eq:quan12}
\Theta(L)-\Theta(0)=-\theta, \ {\rm or}\  , \Theta(x)=-\frac{\theta}{L}x,  
\end{equation}
gives us a $\theta$-independent boundary condition, $\Psi(L)=\Psi(0)$
as in Sec. \ref{sec:a-sing}.  Moreover, a direct substitution shows
that $\Psi(x)$ is the eigenfunction of a transformed
Hamiltonian\footnote{Under a unitary transformation $|\psi'\rangle =U
  |\psi\rangle$, the average of an operator $A$  remain the same so
  that $\langle\psi|A|\psi\rangle=\langle\psi'|A'|\psi'\rangle=\langle\psi'|U^{\dagger}AU|\psi'\rangle
$, identifying the transformed operator $A'=U^{\dagger}AU$.}   but with the same
energy,
\begin{equation}
  \label{eq:quan9}
H_{\theta}=e^{i\Theta(x)}H e^{-i\Theta(x)}=\frac{1}{2m} \left(p+\frac{\hbar\theta}{L}\right)^2,\quad
{\rm and} \quad 
 H_{\theta} \Psi_n(x)=E_n \Psi_n(x),
\end{equation}
where $E$ is given by Eq. (\ref{eq:quan7})
The quantum problem turns out to be equivalent to a
charged particle  in a magnetic vector potential, but with a
$\theta$-{\it independent}  periodic
boundary condition for the wave function.  

Suppose there is a thin solenoid of radius $b$ carrying a magnetic
field $B$ at the center of the ring in a direction perpendicular to
the plane of the ring . There is no magnetic field on the ring but
there exists a vector potential
\begin{equation}
  \label{eq:quan20}
A=\frac{B\pi b^2}{2\pi r},  
\end{equation}
on a circle of radius $r$ in the angular direction.  The Hamiltonian
of a particle of charge $q$ is then $H_{mag}=\frac{1}{2m}(p- qA/c)^2$,
$c$ being the velocity of light.  Comparing this form with Eq.
(\ref{eq:quan9}), we see that $\theta=2\pi \frac{\Phi}{\Phi_0}$, where
$\phi_0=2\pi\hbar q/c$, the standard flux quantum if the charge $q$ is
the electronic charge $e$.

\begin{figure}[htbp]
\begin{center}
\includegraphics[scale=0.51]{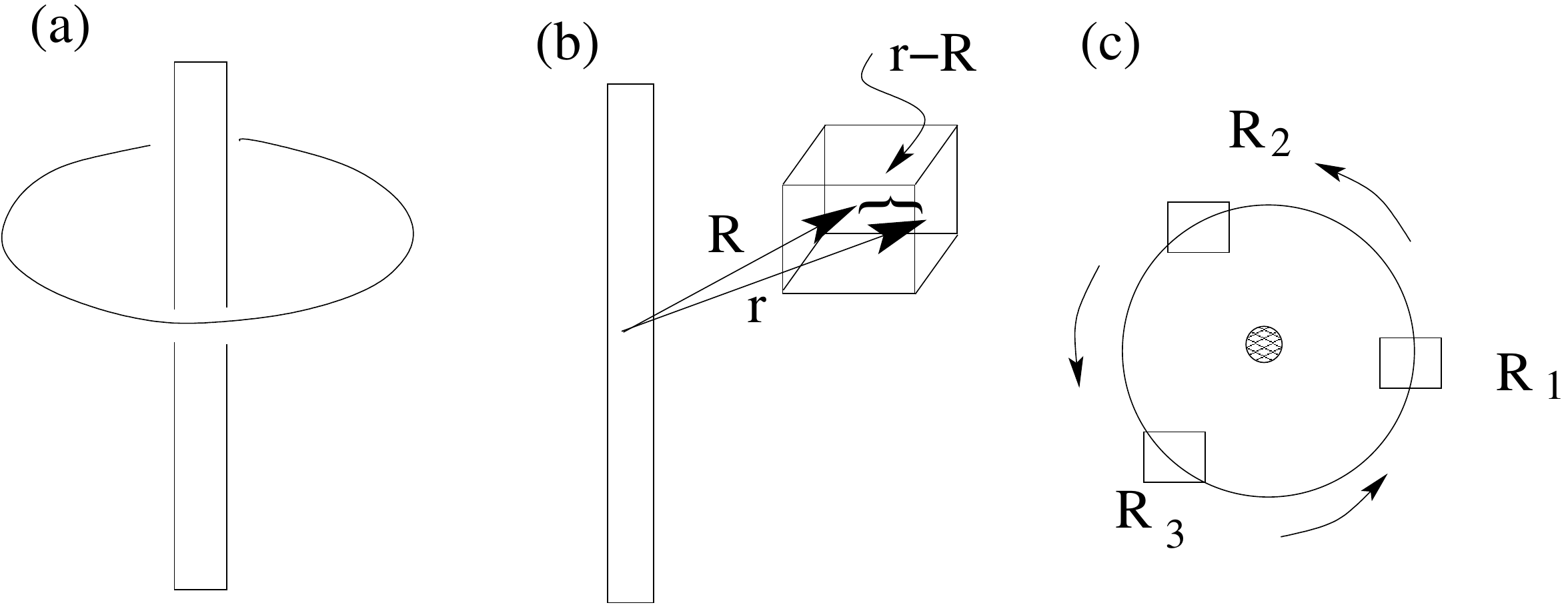}
\end{center}
\caption{(a) A particle on a ring threaded by a flux line through
  origin. (b) A box, confining the particle, centered on ${\bf R}$ on
  the circle. The position of the particle inside the box is ${\bf
    r}$. (c) The boxed is taken around the full circle adiabatically 
in a continuous
  manner, or, say, in three steps so as to enclose the flux.
}\label{fig:quan1}
\end{figure}

It needs to be recognized at this point that the circular ring per se
is not special here; geometry does not matter.  The analysis is valid
for a two dimensional plane threaded by an impenetrable thin flux
line, which is like a hole (See Prob. 2.13).  Phase $\theta$ is
independent of path, allowing us to claim that it is topological in
nature.  This phase is known as the Aharonov-Bohm phase.

\subsection{Topological/Geometrical phase}
\label{sec:berrys-phase}
The extra phase we derived in the preceding case is not the only
one.   Such phases occur in many situations, classical or quantum.  For
waves in classical physics, such a phase is called the
Pancharatnam phase while in quantum mechanics it is Berry's phase.  
As an angle it can be found in classical problems as Hannay angle,
writhes in DNA and so.  Often the angle may be geometric in origin, not
necessarily topological.  What it means is the the angle one gets may
depend on the path chosen unlike the path independence of a
topological quantity.  In this respect it is important to distinguish
between a geometrical phase and a topological phase.

\subsubsection{ Berry's phase}
We established the equivalence of two QM problems, viz.,  
\begin{enumerate}
\item  a charged  particle on a
ring enclosing a constant flux, with {\it single valued} wave function, 
\item  a particle on a ring in zero flux  but with a {\it multivalued}
wave function.
\end{enumerate}
This equivalence raises the following  question,
  \begin{quotation}
 What is the analogue of the  extra phase (case 2)
responsible for the multivaluedness    in the singlevalued version of
case 1?    
  \end{quotation}
The answer lies  in Berry's phase.

\subsubsection{Phase -- an Angle: two formulas}
\label{sec:angle:-formula}
In order to see the emergence of the phase, let us take a wavepacket
localized on the ring.  One way to achieve this is to enclose the
particle in a small box centered on a point on the ring, and then move
the box around the circle.  Let the box position (centre of the box)
be ${\bf R}$ on the ring.  The particle is now described by
\begin{equation}
  \label{eq:quan13}
  \left[ \frac{1}{2m}  \left(p-\frac{qA({\bf r})}{c}\right)^2   + V({\bf r}-{\bf R})\right ]\chi({\bf r},{\bf R})
= {\cal E}\chi({\bf r},{\bf R}),
\end{equation}
where $A({\bf r})$ is given by Eq. (\ref{eq:quan20}). 
Within the finite-sized box the wavefunction  is single valued.   

Now  this box is taken  adiabatically around the circle through a complete turn  as
shown in Fig \ref{fig:quan1}.  We assume that the wave function
changes very slowly, though, in fact, we shall see that the speed doe
not matter.   For two positions ${\bf R}_1, {\bf
  R}_2$, the phase difference is defined as
\begin{equation}
  \label{eq:quan15}
  \exp({- i\phi_{12}})=
\frac{\langle \chi({\bf r},{\bf R}_1)|\chi({\bf r},{\bf R}_2)\rangle}{|\langle \chi({\bf r},{\bf R}_1)|\chi({\bf r},{\bf
    R}_2)\rangle|},\;\; {\rm or\;\; } \phi_{12}=- {\rm Im}\; \ln {\langle \chi({\bf r},{\bf R}_1)|\chi({\bf r},{\bf R}_2)\rangle}.
\end{equation}
This phase between two points is somewhat arbitrary because it can be
changed by a gauge
transformation at any one of the points ${\bf R}_1$ or ${\bf R}_2$.
In spite of this arbitrariness, it can be combined
with the phases from the remaining steps to define an overall phase as
 (${\bf r}$ is suppressed in the notation)
\begin{equation}
  \label{eq:quan16}
  \gamma= - {\rm Im} \ln \left ( {\langle \chi({\bf r},{\bf R}_1)|\chi({\bf
      r},{\bf R}_2)\rangle}{\langle \chi({\bf r},{\bf R}_2)|\chi({\bf
      r},{\bf R}_3)\rangle}.{\langle \chi({\bf r},{\bf R}_3)|\chi({\bf
      r},{\bf R}_1)\rangle}\right)
\end{equation}
Note that the arbitrariness of phases at the intermediate points get
cancelled in the product.  Consequently the total phase $\gamma$
modulo $2\pi$ is a phase that cannot be removed by a gauge
transformation, and, as a gauge invariant quantity, must have physical
consequences.  This phase is an example of what is called {\it Berry's phase}.

 We may take a continuum limit where the closed loop is traversed by
 infinite number steps, with ${\bf R}_{i+1}={\bf R}_i+d{\bf R}$.
 Using continuity, $\chi{\bf R}+d{\bf R})=\chi({\bf R}) + \nabla_R
 \chi({\bf R})\cdot d{\bf R}$, and then expanding the logarithm, the
 phase factor can be written as an integral (taking the wavefunctions
 to be normalized)
\begin{equation}
  \label{eq:quan17}
  \gamma= i \oint d{\bf R} \cdot \langle \chi({\bf R})|\nabla_R \chi({\bf R})\rangle, 
\end{equation}
taking the wavefunctions to be normalized.

Whenever a Hamiltonian depends on a parameter (no quantum evolution of
this parameter) and the parameter goes through a cyclic path in the
parameter space, the parameter-dependent wavefunction develops a phase
given by either Eq. (\ref{eq:quan16}) or Eq. (\ref{eq:quan17}).  In
numerical computation where the eigenfunctions are determined
numerically -- and therefore with unknown phases -- Eq.
(\ref{eq:quan16}) is preferable because, by construction, the
intermediate unknown phases cancel out.  In many analytical approaches
for which a continuous wavefunction is known, Eq. (\ref{eq:quan17}) is
useful.

\subsubsection{Berry's phase and the Aharonov-Bohm phase}
\label{sec:berrys-phase-ahar}
We now show that the unavoidable Berry's phase in the formulation
with a magnetic flux is the extra phase in the multivalued
formulation.

Eq. (\ref{eq:quan13}) can be solved by a gauge transformation
$\chi= e^{ig(x)} \widetilde{\chi}$ so that 
\begin{equation}
  \label{eq:quan14}
  \left[ \frac{ p^2 }{2m}  + V({\bf r}-{\bf R})\right ]{\widetilde{\chi}}({\bf r}-{\bf R})
= {\cal E}{\widetilde{\chi}}({\bf r}-{\bf R}),\quad {\rm with}\quad g(x)=\frac{q}{\hbar c} \int_{\bf R}^{\bf{r}} A({\bf r}') d{\bf  r}.
\end{equation}
This is very similar to what we did earlier in Sec B2, but here the
wave function ${\widetilde{\chi}}$ remains singlevalued, mainly because
the interior of the box is a simply-connected region.

To use Eq. (\ref{eq:quan17}), we need $\nabla_R \chi({\bf r},{\bf R})$
which can be written as\footnote{Note that $\nabla_R f({\bf r}-{\bf
    R}))=-\nabla_r f({\bf r}-{\bf R}))$}
\begin{equation}
  \nabla_R \chi({\bf r},{\bf R})=
  \nabla_R \{e^{ig} {\widetilde{\chi}}({\bf r}-{\bf R})\}
=-i \frac{q}{\hbar c} A({\bf R}) \chi - e^{ig} \nabla_r {\widetilde{\chi}}({\bf r}-{\bf R})\}, \label{eq:quan18}
\end{equation}
so that (taking normalized $\widetilde{\chi}$)
\begin{equation}
   \label{eq:3a}
  \langle  \chi({\bf R})|\nabla_R \chi({\bf R})\rangle
= -i  \frac{q}{\hbar c}    A({\bf R}) -   \langle  {\widetilde{\chi}}({\bf
  r}-{\bf R})|\nabla_r {\widetilde{\chi}}({\bf r}-{\bf R})\rangle.
\end{equation}
As the average momentum in the localized state in the box is zero, we
obtain the overall phase on taking the box around the loop once
\begin{equation}
  \label{eq:quan19}
  \gamma=\oint \frac{q}{\hbar c} A({\bf R})\cdot d{\bf
    R}=\frac{q}{\hbar c} \ \int d{\bf S} \cdot {\bf{\nabla}}_R\times A({\bf
    R})=\frac{q}{\hbar c}\Phi = 2\pi \; \frac{\Phi}{\Phi_0}=\theta,
\end{equation}
precisely the same angle  we saw in Eq. (\ref{eq:quan12a}).
The line of arguments here indicates that the angle is independent of
the size and shape of the loop in the plane so long it encloses the
origin once.  The answer is ultimately determined by the number of
times (=1) the flux tube pierces the surface used in the surface
integral.   The Aharonov-Bohm phase, viewed as Berry's phase, is
therefore topological in essence.  

To summarize, in a topologically nontrivial space, we may either use
multivalued wavefunctions or use a gauge transformation to a magnetic
field like problem with singlevalued wavefunction that admits a
geometric phase, Berry's phase.  A generalization, without proof, is
that a topological phase (Berry's phase) occurs\footnote{With complex
  wavefunctions, Berry's phase may occur in simplyconnected space
  too.}  if (i) a parameter, $R$, defined on a multiplyconnected
space, is taken around a nontrivial loop, and (ii) the space of the
wavefunctions remains the same as the parameter is changed, i.e. the
Hilbert space is independent of $R$.

Another lesson we learnt from this is that a hole or impenetrable
region can be replaced by a vector potential or a gauge term where the
$\theta$ parameter determines the effective flux.  Such assignments of
flux tubes become useful in many situations like anyons.

\subsection{Generalization -- Connection, curvature}
\label{sec:a.4:-gener-conn}

Proper settings for generalization of the above ideas require the
concepts of fibre bundles and  differential forms.  Without going into
those, let us define the terms -- with a little abuse of
definitions -- connection and curvature.  The vector potential ${\bf A}$
we defined is called Berry connection, though actually it should be 
1-form meaning something like ${\bf A}\cdot d{\bf r}$.  The
``magnetic field'' is the curvature.  Again, curvature should be a
2-form meaning objects like ${\bf B}\cdot d{\bf s}$ with $d{\bf s}$ as
the area element.  In our convention, the integrals will have
the infinitesimals $dr, ds$ explicitly.

The general formulas for connection and curvature for a state, 
labelled  by 
$m$, and, dependent on a set of parameters $R_{\mu}$, are
\begin{eqnarray}
  \label{eq:16}
  {\rm Berry\ connection:}\;\;A_{\mu}&=&i \left\langle m
    R\left|\partial_{\mu}\right|m R\right\rangle,\quad \left(\partial_{\mu}\equiv\frac{\partial\ }{\partial
    R_{\mu}}\right),\\
  {\rm Berry\ curvature: }\;\;\Omega_{\mu\nu}&=&\partial_{\mu}A_{\nu} -  \partial_{\nu} A_{\mu}.\label{eq:22}
\end{eqnarray}
In three dimensions, i.e. if $R$ has three components,  the
curvature tensor can be written as a vector, 
\begin{equation}
  \label{eq:19}
  B_{\lambda}=\frac{1}{2}\; i\;  \varepsilon_{\lambda\mu\nu} \; \left(\partial_{R_{\mu}}\langle m R|\right) \;\left(\partial_{\nu}|m R\rangle\right),
\end{equation}
where $\varepsilon_{\lambda\mu\nu}$ is the usual antisymmetric tensor.  The
state index $m$ has been omitted from the notation of $A,\Omega$.  The
connection is like the vector potential, while, from Eq. (\ref{eq:19}),
${\bf B}={\bf{\nabla}}\times {\bf A}$ is like a magnetic field.  For
generality, instead of linking these to electromagnetism, we call
$\Omega_{\mu\nu}$ as the Berry curvature per unit area.  An integral of
the curvature over an open surface $S$ (with boundary) gives the phase
(Berry's phase) associated with the closed loop, the boundary of $S$.

For a given Hamiltonian $H(R)$, every eigenstate will have its own Berry
connection and Berry curvature.  
Defining the eigenvalue equation as  $H(R)|n R\rangle=E_n |n
R\rangle$, with no degeneracy, a straightforward manipulation shows that the Berry
curvature for the $n$th state is (see problem)
\begin{equation}
  \label{eq:23}
  \Omega_{\mu\nu}^{n}=i \sum_{p\neq n} \;  \frac{\langle nR|\partial_{\mu} H(R)|p
      R\rangle \langle pR|\partial_{\nu} H(R)|n
      R\rangle}{[E_p(R)-E_n(R)]^2} - \{\mu\leftrightarrow \nu\}, 
\end{equation}
which has the advantage that the derivatives are now of the
Hamiltonian and not of the wavefunctions. It also follows from the
antisymmetric nature that 
\begin{equation}
  \label{eq:24}
 \sum_n\Omega^n_{\mu\nu}=0. 
\end{equation}
Eq. (\ref{eq:23}) shows that the Berry curvature is large for ``near
degeneracies'' or, equivalently, a large Berry curvature can be taken
as a signal for nearby eigenvalues.  If there is a degeneracy, then
one has to project out that part of the space leading to nonabelian
issues.

The degeneracy points are singular points in the $d$-dimensional
parameter space.  The loops for Berry's phase need to enclose these
singular points for a nonzero value.  The loops actually tell us about
the first Homotopy group of the allowed part of the space as the loop
is not allowed to go through the singular point.  Loops in
2-dimensions enclose point defects, in 3-dimensions line defects and
so on, so that the singular points in $d$-dimensions must form a
$(d-2)$-dimensional space (or manifold) for the first homotopy group
of the allowed space is nontrivial.  This restriction provides a quick
check when not to expect any topological phase.

\subsection{Chern, Gauss-Bonnet}
\label{sec:4:-chern-gauss}

In the examples we consider, the space of the parameter $R$ is an even
dimensional closed surface ${\cal M}$.  For concreteness, let ${\cal
  M}$ be a two-dimensional closed orientable surface that can be
embedded in three dimensions.  By orientable we mean at at each point
we can define a unique normal to the surface.  Simple examples are
$S^2,{\mathbb{T}}^2$, etc.  There are now two topological problems in
hand.  One is the topological characterization of ${\cal M}$ and the
other one is that of the map from ${\cal M}$ to the manifold of
wavefunctions.  Two theorems are useful here, (i) the Gauss-Bonnet
theorem involving the geometric curvature of ${\cal M}$ and Chern's
theorem involving the Berry curvature.

Chern's theorem states that the integral of the Berry curvature (a
geometric quantity) over the closed surface is equal to $2\pi$ times
an integer, i.e.,
\begin{equation}
  \label{eq:25}
C_1=  \frac{1}{2\pi} \int_{\cal M}\; {\bf \Omega}\cdot d{\bf S} = n \in
  {\mathbb Z},\qquad {\rm(Chern's\;\; theorem)}. 
\end{equation}
This number $C_1$ is called the {\it first Chern number}.\footnote{The
  $1/(2\pi)$ factor actually comes from a general factor $2/K_d$, where
  $K_d=\frac{2\pi^{d/2}}{\Gamma(d/2)}$ is the volume of $S^{d-1}$ or
  the surface area of a $d$-dimensional sphere,
  which occurs for the theorem for a higher dimensional closed
  surface.   This is applicable to the Gauss-Bonnet theorem too. For Eq.
  (\ref{eq:25}), put $d=3$.} Two mappings (or states in this case)
with different first Chern numbers cannot be continuously deformed
into each other.  In other words, to go from one to the other by
tuning some parameter (not $R$), there has to be a topology change at
some special value of the parameter. This corresponds to a phase
transition or a quantum critical point.  Of these, $C_1=0$ is called a
trivial phase while $C_1\neq 0$ are nontrivial topological
phases.\footnote{Warning: Phase here means a state of the system like
  liquid, gas etc, and not the phase of a wavefunction!}  As an
analogy one may refer to Fig. \ref{fig:12}c, where the oscillatory and
the circular motions are separated by the special figure 8 space.
 
The Chern number is a topological characteristic of the manifold of
the energy eigenstate defined on ${\cal M}$ and is not just a
topological property of ${\cal M}$.  The topological characteristic of
${\cal M}$ comes from the Gauss-Bonnet theorem,  which for a closed
two dimensional surface states that the surface integral of the Gaussian
curvature is a topological quantity, viz., 
\begin{equation}
  \label{eq:26}
\frac{1}{2\pi}\; \int K dS=\chi=2(1-g),  ({\rm Gauss-Bonnet\;\; theorem})  
\end{equation}
 where $\chi$ is the Euler characteristic and 
$g$ is the genus of the surface.  Both $\chi$ and $g$ are topological properties of
any surface.  For a sphere of radius $r$, the Gaussian curvature
(=product of the two principal curvatures at a point) is uniform,
$K=1/r^2$, and its genus $g=0$.   Eq. (\ref{eq:26}) is then obviously
satisfied. 

\subsection{Classical context: geometric phase}
\label{sec:class-cont-geom}
To show that the angle is not just a quantum mechanical issue, let us take a
classical example.

\begin{figure}[htbp]
\begin{center}
\includegraphics[scale=0.5,clip]{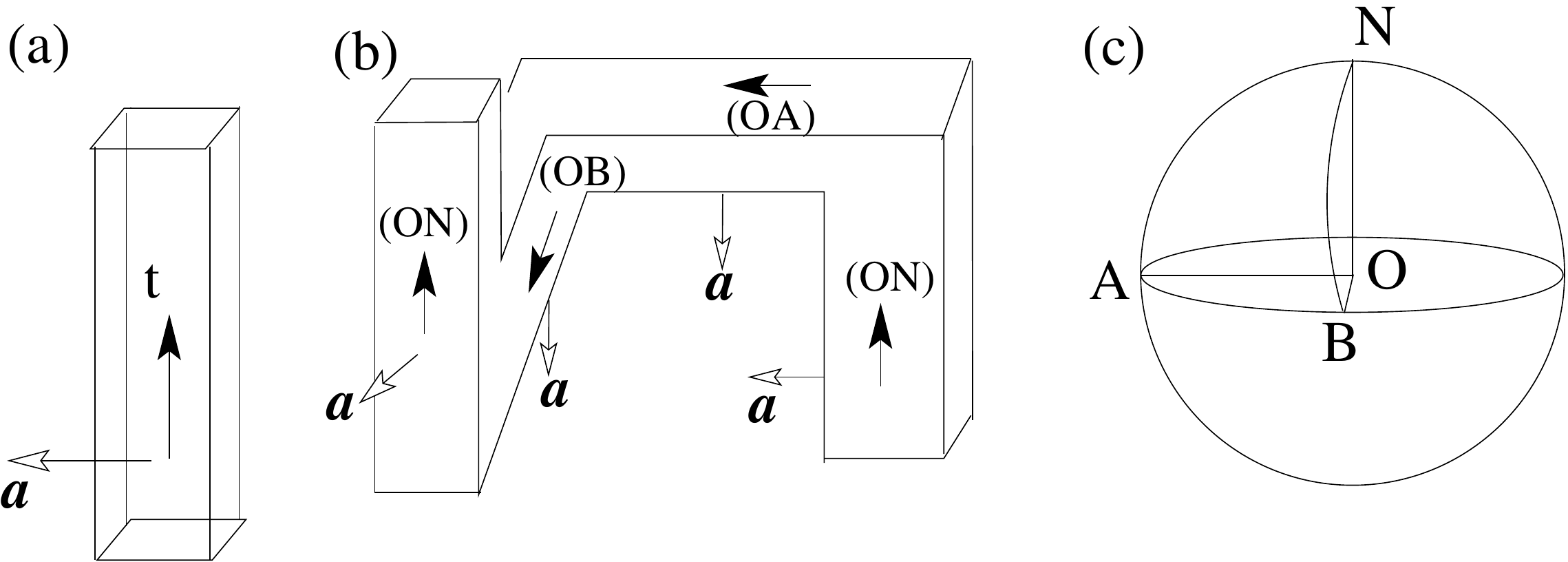}
\end{center}
\caption{(a) A long bar of rectangular cross section.  Unit vector
  ${\bf t}$ is along the axis, and ${\bf n}$ perpendicular to one
  side.  (b) The bar is bent into a twisted form.  The two vectors at
  intermediate positions are as shown.  A change in the orientation
  of ${\bf n}$ by $\pi/2$ is visible as ${\bf t}$ returns to its
  original orientation.  (c) $S^2$ as the space for vector ${\bf t}$.
  As one moves along the bar, ${\bf t}$ goes from ON to OA, OA to OB,
  and then back to ON, a closed path on $S^2$.  The solid angle
  subtended by the closed region NABN is $\pi/2$.  In the quantum
  mechanics problem, there is an extra factor of $1/2$ (see Eq. (\ref{eq:18})). }
\label{fig:bar}
\end{figure}

Take a long bar of rectangular cross section so as to identify the
sides easily.  See Fig. \ref{fig:bar}.  Define unit vectors ${\bf t}$
along the axis, and ${\bf a}$ perpendicular to one side.  Orientations
are to be kept fixed locally.  The bar is bent into a twisted form as
in Fig. \ref{fig:bar}b.  The two vectors are monitored along the tube,
keeping their orientations fixed locally.  A $\pi/2$ change in the
orientation of ${\bf a}$ is visible even when ${\bf t}$ gets back to
its original orientation.  To see this change, we note that vector
${\bf t}$, as a unit vector, spans a sphere $S^2$ as in Fig.
\ref{fig:bar}c.  As one moves along the bar, ${\bf t}$ goes from ON to
OA, OA to OB, and then back to ON, a closed path on $S^2$.  The solid
angle subtended by the closed region NABN is $1/8$ of the sphere, i.e.
$\pi/2$ which is the change in orientation of ${\bf n}$.  One may
straighten the bar in Fig. \ref{fig:bar}b to see that there is a twist
though the axis may remain straight.

This is an example of a geometric phase, not necessarily topological,
because the solid angle subtended by a closed loop depends on the
details of the loop.  The rotation of the Foucault pendulum as the
earth rotates under it is another example of this geometric phase.
This particular example and its generalizations are important in
macromolecules like DNA and is called {\it twist}. 

For a classical wave,  polarized light, with ${\bf n}$ as the
polarization direction and ${\bf t}$ as the direction of propagation,
one may see a change  in the direction of polarization.  The angle
appears as a phase there and is the classical analog of Berry's
phase.  The classical phase is called the {\it Pancharatnam phase}.
In classical dynamics, such an angle also occurs and is known as
the {\it Hannay Angle}.  For details on these see Ref. \cite{wilc}

Analogous to this classical example, Berry's phase is generally  a geometrical
phase, and in special situations, like the Aharonov-Bohm case, it becomes a
topological phase.  To repeat, a topological phase is independent of the
details of the path (same for all homotopic loops) while a geometrical
phase is, in general,  dependent on the details of the loop.

\subsection{Examples: Spin-$1/2$ and Quantum two level system}
\label{sec:quantum-two-level}

\subsubsection{Spin-$1/2$ in a magnrtic field}
\label{sec:spin-12-magnrtic}

A counterpart of the problem discussed in Sec.
\ref{sec:class-cont-geom} is a spin $1/2$ in a magnetic field with a
Hamiltonian $H=-{\bf d}.\sigma$, where $\sigma$ is the 3-d vector of
the Pauli spin matrices (See Sec \ref{sec:space-hamilt-two}).  For a
given field ${\bf d}$, there are two eigenstates, $\pm |d|$ with
eigenvectors parallel or antiparallel to the direction of ${\bf d}$.
As a vector, the field of constant magnitude $|d|$ can be in any
direction in 3-dimensions, ${\bf d}=|d|\;\;(\sin \theta \cos\;\phi,
\sin\;\theta \sin\;\phi,\cos\;\theta )$ with $\theta,\phi$ as the
usual polar angles, spanning a sphere.  The relevant space for the
field is $S^2$.  Let us choose the eigenstate for energy $+|d|$,
(compare with Eq. (\ref{eq:7})),
\begin{eqnarray}
  \label{eq:14}
|u \rangle = \left(\begin{array}{c}
          \sin \frac{\theta}{2} e^{-i\phi} \\
          -\cos\frac{\theta}{2} 
          \end{array}\right),
\end{eqnarray}
upto an arbitrary phase factor.  This $|u\rangle$ is a spinor, but,
is, otherwise, fixed by the direction of the field ${\bf d}$.  If the
magnetic field is now rotated, it forms a closed loop on the sphere.
To get the phase acquired by the wavefunction as given by Eq.
(\ref{eq:quan19}), embed the sphere in three dimensions \footnote{We
  are embedding to take advantage of the vector notation.  In
  spherical polar coordinates, for any vector ${\bf A}=A_r {\hat
    r}+A_{\theta} {\hat{\theta}}+A_{\phi} {\hat{\phi}}$,
  \begin{eqnarray*}
    {\bf \nabla}=\hat{r}
\frac{\partial \ }{\partial r} + \hat{\theta}\ \frac{1}{r}\;
\frac{\partial\ }{\partial \theta}+
\hat{\phi}\frac{1}{r\sin\theta}\;\frac{\partial\ }{\partial \phi},
\quad {\bf \nabla}\times{\bf A}=\left( \begin{array}{ccc}
        (r^2\sin\theta)^{-1} {\hat{r}}&
        (r\sin\theta)^{-1}{\hat{\theta}}& r^{-1} {\hat{\phi}}\\
          \partial/\partial r&
          \partial/\partial\theta&\partial/\partial\phi\\
          A_r& rA_{\theta}& r \sin\theta A_{\phi}
\end{array}\right).
  \end{eqnarray*}
} 
to obtain
\begin{equation}
   \label{eq:17}
{\bf A}=\langle u|i{\bf \nabla_d}|u\rangle
=\frac{\sin^2\frac{\theta}{2}}{|d|\;\sin\theta} \hat{\phi}=\frac{1}{2}\; \frac{1-\cos\theta}{|d|\;\sin\theta} \hat{\phi},\ {\rm  with}\  {\bf \Omega}=
{\bf \nabla}\times {\bf A}=-\frac{1}{2} \frac{{\bf d}}{|d|^3},   
\end{equation} 
in the radial direction.  The line integral may be converted to a
surface integral with the help of Stokes' theorem, by choosing the
enclosed part of the sphere as the relevant surface.  As the area vector is
radial, we see 
\begin{equation}
  \label{eq:18}
\gamma=\int {\bf \Omega}\cdot d{\bf s}=-\frac{1}{2}\int
d\omega=-\frac{1}{2} \;\mho_c,
\end{equation}
$d\omega,$ being the angular part of the spherical integral and
$\mho_c$ is the solid angle formed by the closed loop.  The similarity
with the classical case of Fig \ref{fig:bar} is to be recognized,
except for the factor of $1/2$ which is purely a quantum mechanical
contribution.  A solid angle is a geometrical quantity, dependent on
the closed path, and so the phase here, unlike the Aharonov-Bohm
phase, is not topological.  The form of the ``magnetic field'' ${\bf
  B}$ shows that there is a singularity at $d=0$, the degeneracy point
and the functional form of ${\bf \Omega}$ satisfies ${\bf \nabla}\cdot
{\bf \Omega}=4\pi \; \delta({\bf r})$ with $g=-1/2$ and $\delta({\bf
  r})$ as the Dirac $\delta$-function.  With ${\bf \Omega}$ as a
``magnetic field'', it looks like there is a magnetic monopole
$g=-1/2$ at the center, and the Berry phase is the flux due to this
monopole through the loop.

The magnetic monopole interpretation helps us in identifying a
topological invariant, using the equivalent of the Gauss theorem.  An
integration over the whole sphere is the flux through the closed
surface and it counts the number of monopoles
enclosed by the sphere.  In other words
\begin{equation}
  \label{eq:15}
 C_1\equiv  \frac{1}{2\pi}\ \int_{S^2}  {\bf \Omega}\cdot d{\bf s}=n\in \mathbb{Z}.
\end{equation}
This is  the {\it first Chern number} as defined in Eq. (\ref{eq:25}).

It is important to see from a different view point why $C_1\neq 0$.
Let us divide the surface integral into parts at the equator.  The
closed loop integral $\oint {\bf A}\cdot d{\bf l}$ can be evaluated
with the help of Stokes theorem with either the upper hemisphere
containing the North pole or the lower hemisphere containing the South
pole, provided ${\bf A}$ can be defined uniquely on the chosen surface
and the equator.  Since Eq. (\ref{eq:14}) is valid for the part of the
sphere that encloses the North pole, we get, for the upper hemisphere
(uh), $\oint {\bf A}\cdot d{\bf l}=\int_{\rm uh} d{\bf s}\cdot {\bf
  \Omega}$ which evaluates to $+\pi$.  However, this choice of
$|u\rangle $ cannot be used for the lower hemisphere (lh), as the
spinor, Eq. (\ref{eq:14}), has undefined phase at the South pole
($\theta=\pi$) where $|u\rangle=(e^{i\phi}\ 0)^T\equiv (1 \;\; 0)^T$.
A possible choice (``choice of gauge'') is $|u_{lh}\rangle=( \sin
(\theta/2), -\cos (\theta/2) e^{i\phi})^T$ which is now defined
everywhere on the sphere except the North pole ($\theta=0$).  With
this choice ${\bf A}=- \frac{1}{2}\;
\frac{1+\cos\theta}{|d|\;\sin\theta} \hat{\phi}$, though the curvature
($\Omega$) remains the same as Eq. (\ref{eq:17}).  A direct line
integral over the equator ($\theta=\pi/2$) gives $-\pi=\pi -2\pi$.  In
other words it differs from the upper hemisphere result by $-2\pi$
which does not matter as an angle.  If Stokes theorem is used, $\oint
{\bf A}\cdot d{\bf l}=-\int_{lh} {\bf \Omega}\cdot d{\bf s} $ with a
minus sign coming from the direction rule of the theorem.  Since it is
the same line integral, We must have $(\int_{uh}+\int_{lh}) {\bf
  \Omega}\cdot d{\bf s}=0$ upto $2\pi n, n\in {\mathbb{Z}}$.  In this
particular case, we find $-2\pi$ whose origin lies in the
nonuniformity of the guage choice.  If a single gauge choice can be
done over the whole surface, then the surface integral would have been
zero.  Thus the zero Chern number corresponds to the trivial case,
while a nonzero Chern number tells us that more than one map is needed
to cover the whole closed surface. In this case we need two.

For the $S^2$ case, the state is given by $-{\hat{n}}$ which is also
the normal to the surface. The Berry curvature, Eq. (\ref{eq:22}) is given by
\begin{equation}
  \label{eq:21}
  \Omega_{lm}= -\frac{1}{2}{\hat{n}}\cdot
\left(\frac{\partial {\hat{n}}}{\partial R_l}\times \frac{\partial
    {\hat{n}} }{\partial R_m}\right),
\end{equation}
where the derivatives are now the angular derivatives, and it is same
as ${\bf B}$ in Eq. (\ref{eq:17}).  The Chern number can therefore be
written as
\begin{equation}
  \label{eq:20}
 C_1=-\frac{1}{4\pi}\ \int_{S^2}\; ds\;\;  {\hat{n}}\cdot
\left(\frac{\partial {\hat{n}}}{\partial \theta}\times \frac{1}{\sin\theta}\frac{\partial
    {\hat{n}} }{\partial \phi}\right)=-1.
\end{equation}

The analysis for Berry's phase can be done for the other state
$|v\rangle$ with eigenvalue $-|d|$.  In that case the monopole will be
of charge $+1/2$ so that the total of all the states is zero,
consistent with Eq. (\ref{eq:24}).

\subsubsection{Case of two bands: Chern insulators}
\label{sec:d.-case-two}

The results of the two level system in the previous section finds a
ready use in a two band system.   This  situation  arises for a band
insulator where we may focus on the last occupied band and the next
unoccupied band.

\begin{figure}[htbp]
\begin{center}
\includegraphics[scale=0.5,clip]{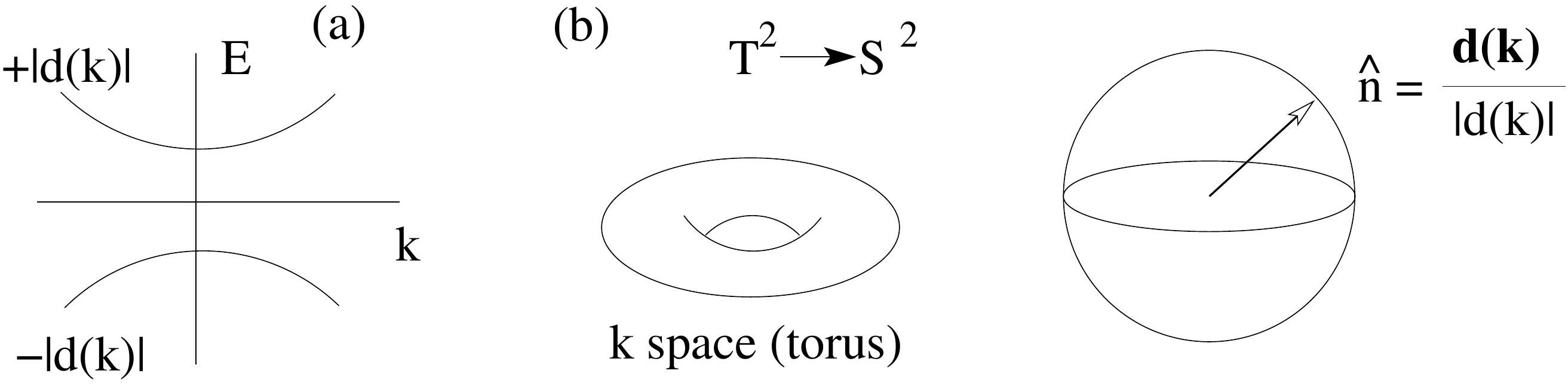}
\end{center}
\caption{(a) Schematic diagram of two bands vs k.  The band energies
are $\pm |d({\bf k})|$. (b) In two dimensions the k-space is a torus
${\mathbb{T}}^2$ while  unit vector ${\hat{n}}({\bf k})= {\bf d}({\bf
      k})/|d({\bf k})|$ lies on a sphere. This gives a
    ${\mathbb{T}}^2\to S^2$ map.  }
\label{fig:band}
\end{figure}

The bands are described by the pseudomomentum ${\bf k}$ in the first
Brillouin zone.  The Hamiltonian can be written as $H=-{\bf d(k)}\cdot
{\bf \sigma}$, with the bands given by $\pm |d({\bf k})|$.  For a two
dimensional problem, the $k$-space (or the first Brillouin zone) is a
torus ${\mathbb{T}}^2$, while unit vector ${\hat{n}}({\bf k})= {\bf
  d}({\bf k})/|d({\bf k})|$ maps out a sphere $S^2$.  It may be
visualized as a three component vector ${\bf n}$ attached to every
point of the Brillouin zone torus.  One may compare with the
Heisenberg magnet example of Sec. \ref{sec:magnets}, where we looked
at the arrangements of three-vectors in Eucildean space.    Here,
instead of the Euclidean space, we now have a torus.   All possible
insulators can now be characterized by the ``spin arrangements'' on
the torus.   In the real space case of magnets, we found
$\pi_1(S^2)=0$, and so there is no topological distinction among the 
spin configurations in space.   In the present case, the mapping is
nontrivial, and that tells us that all insulators are not
topologically equivalent.  In other words, there are topologically
inequivalent classes of insulators, identified by the arrangements of
the gap vector ${\bf d}$ on the Brillouin Zone, which is a torus. 
The
nontrivial ones are called topological insulators.

  As we move along the torus, the
${\bf k}$-vector changes, and it acts as the parameter for Berry's
phase or equivalently the curvature.  The topological invariant is
then the first Chern number that tells us how many times (with sign)
vector ${\hat{n}}$ goes around the sphere as one traverses the torus.
This number comes from the Chern formula
\begin{equation}
  \label{eq:27}
  C_1=\frac{1}{4\pi}\ \int\int d^2k\;\;  {\hat{n}}\cdot
\left(\frac{\partial {\hat{n}}}{\partial k_x}\times \frac{\partial
    {\hat{n}} }{\partial k_y}\right).
\end{equation}
We have talked about the homotopy groups which allows one to explore a
space by spheres $S^n$ for various integer $n$.  But, instead of
spheres, we may explore the space by tori as well.  For the problem in
hand, we are classifying the configurations of 3-component vectors on
the torus.  This is described by homotopy[${\mathbb{T}}^2,S^2$] which
is known to be ${\mathbb{Z}}$.  The first Chern number is a concrete
way of getting this integer for a particular case.

Let us choose an example,
\begin{equation}
  \label{eq:33}
{\bf d}({\bf k})= (\sin k_x, \sin k_y, r+\cos k_x+\cos k_y)).    
\end{equation}
If $|r|$ is very large, ${\hat n}$ is nearly
along $(0,0,\pm 1)$, and so does not wind completely the sphere.  The
Chern number is zero as can be checked from Eq. (\ref{eq:27}) for
$|r|>2$.  One can easily check that $C_1=1$ if $-2<r<0$, while
$C_1=-1$ if $0<r<2$.   There is topology change but the change is not  
obvious from the band structure. 
The Chern number is not defined at $r=0,\pm 2$,
the transition points.  Exactly at $r=-2$, we see $|d|=0$ at
$k_x=k_y=0$; this  means the gap closes (no longer an insulator) at
one point.  For $r=2$, the gap closing occurs at $(\pi,\pi)$, while for
$r=0$, it occurs at $(\pi,0)$ and $(0,\pi)$.  The mapping to a sphere
fails when a gap closes, and, consequently, a gap closing is important
for a change of topology.

Band insulators with first Chern number=0 are called trivial
insulators while those with $C_1\neq 0$ are called Chern insulators.\cite{pt}.

\begin{problem}
If the Hamiltonian for a two level system is real, i.e. $d_y=0$, show
that Berry's phase can be $0$ or $\pi$.
\end{problem}
\begin{problem}
  General spin case: A spin-$J$ particle is in a magnetic field with
  Hamiltonian $H=-{\bf d}\cdot {\bf J}$.  Consider any of the
  eigenstates, say, state $|J\; M_J\rangle$ which is an eigenstate of
  the Hamiltonian.  Calulate the Berry phase when ${\bf d}$ rotates as
  in Sec. \ref{sec:quantum-two-level}.  It is easier to use Eq.
  (\ref{eq:23}).  The answer is $-M_J \mho_c$, generalizing Eq.
  (\ref{eq:18}).
\end{problem}

\begin{problem}
  Take the problem of a particle on a ring of radius 1, in presence of
  a magnetic flux threading the ring.  This is the problem discussed
  above but there is now a potential $V(x)=v\delta(x)$ on the ring so
  that the Hamiltonian is
\begin{equation}
  \label{eq:5}
  H=\frac{1}{2m} (p+\theta)^2 + v \delta(x),
\end{equation}
under periodic boundary condition  $\psi(x)=\psi(x+2\pi)$.  Use topological
arguments and necessary gauge transformations to identify this problem
as  the Dirac comb problem (one dimensional Kronig Penny model).  Use
Bloch's theorem to show that  $\theta$ plays the role of the quasi-momentum. 
\end{problem}

\begin{problem}
For a one dimensional model, the Brillouin zone is a circle.  The two
band problem then corresponds to  a mapping $S^1\to S^2$.
  Discuss the nature of this mapping.

 Discuss the general $d$-dimensional case,  $T^d\to S^2$.
\end{problem}
 
\begin{problem}
Complete the calculations for the Berry phase, Berry curvature and the
monopole, counterparts of Eqs. (\ref{eq:17})-~(\ref{eq:20}),  for the
eigenstate with eigenvalue $-|d|$.
\end{problem}
\begin{problem}
Take the example of Eq. (\ref{eq:33}) for three different values of
$r$, $r=-1,1,3$.  Draw the energy bands (3d plot) against
$k_x,k_y$. Separately, map out the region on a sphere traced out by ${\bf
  n}={\bf d}({\bf k})/|d({\bf k})|$ as ${\bf k}$ is taken over the
Brillouin zone.  
\end{problem}

\begin{problem}
  The two bands in Sec. \ref{sec:d.-case-two} are taken to have a
  special symmetry so that the midpoint of the gap is independent of
  ${\bf k}$.  In general, the form of the Hamiltonian should be $H=
  -{\bf d(k)}\cdot {\bf \sigma}+d_0({\bf k}) {\bf I}$, so that the
  space for the Hamiltonian is 4-dimensional.  Show that the arguments
  of that section are not affected by $d_0({\bf k})$.  In other words,
  it is justified to consider a 3-dimensional subspace.
\end{problem}

\begin{problem}
  For the Aharonov-Bohm geometry, we saw the importance of winding
  around the hole.  Consider the free particle case in a plane with a
  hole at origin. By a transformation $t=i\tau$ (imaginary time
  transformation), the Schr\"odinger equation can be converted to a
   diffusion equation which describes a Brownian particle in continuum.
  The winding of the Brownian particle around the hole can be measured
  by making the angle $\theta$ a real variable (refer to Fig.
  \ref{fig:pend1}b,c).  Show that for such a Brownian particle in a
  plane with a hole at origin, the probability distribution for
  winding angle $\theta(t)$ for large $t$ is
\begin{equation}
  \label{eq:29}
P(n)=\frac{1}{\pi}\; \frac{1}{1+x^2},\quad {\rm where}\;
x=\frac{2\theta(t)}{\ln t}. \quad {\rm (Spitzer\  law)}
\end{equation}
The infinite variance of this distribution is because of the large
number of very small windings a particle can do around the hole.
\end{problem}

\section{DNA}
\label{sec:dna}

A situation where the topology of two circles is  needed is DNA.  We
saw the importance of two circles $S^1$ in Fig. \ref{fig:pend3}.  DNA
involves a different type of topological problem.  

\begin{figure}[htbp]
  \begin{center}
 \includegraphics[scale=0.51,clip]{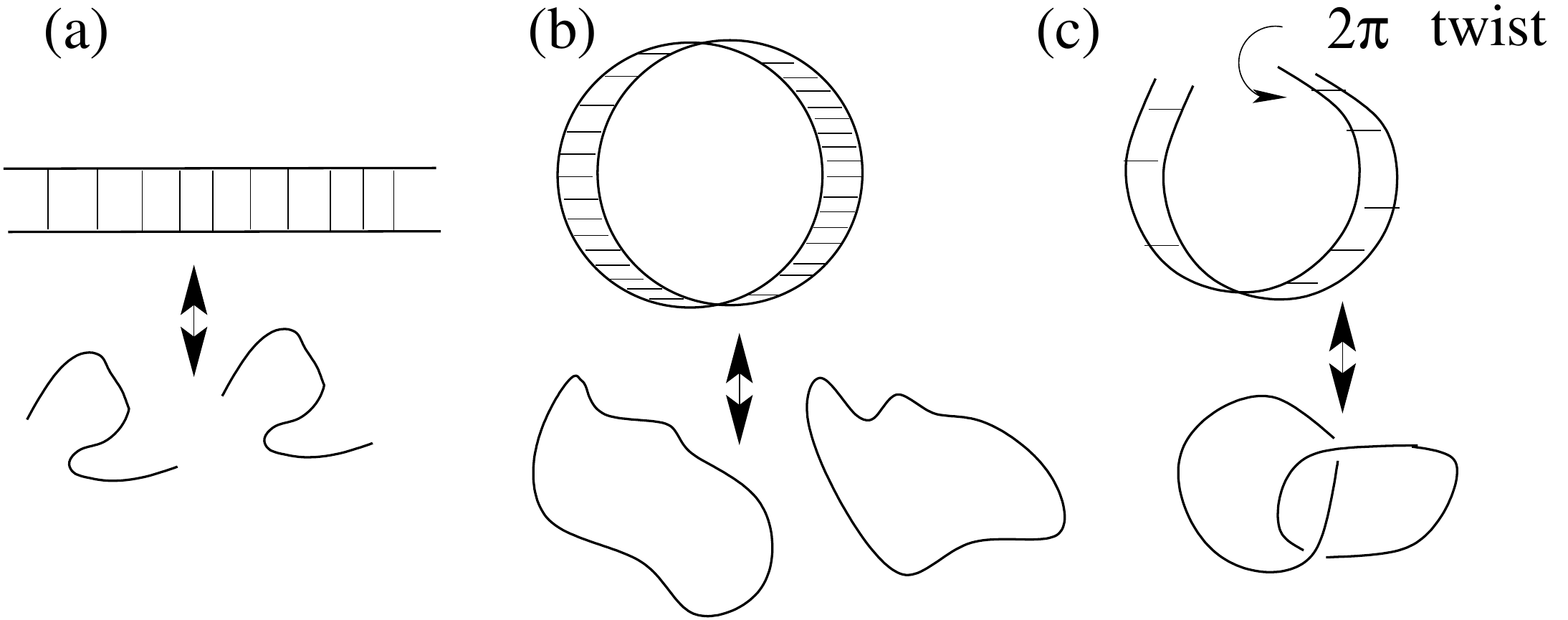} 
\end{center}
\caption{DNA. (a) open DNA with $N$ base pairs.  On melting it gives
  two separate strands. (b) A circular DNA on melting gives two
  separate rings. (c) If a $2\pi$ twist is given before forming the
  closed loop, then on melting there are again two nonpaired rings but
  now topologically linked.  Helicity of the DNA is ignored here.  }
  \label{fig:dna}
\end{figure}
Double stranded DNA consists of two chains, called strands, connected
together by base pairings.  Suppose there are $N$ base pairs, making
the strand length proportional to $N$.  See Fig \ref{fig:dna}.  
In the completely bound state, the energy is $E=-N\epsilon$, where
$-\epsilon$ is the hydrogen bond energy and $N$ is large. If the double stranded DNA
has an entropy $s_b$ per base pair, then the bound state free energy
is given by $F_b= -N\epsilon -TNs_b$  at temperature $T$.   On the other hand if all the
hydrogen bonds are broken, then there are two nonpaired separate
chains, each with entropy $s_0$ per base.  Since there is no energetic
contribution, the free energy of the unbound state $F_u= -2NT s_0$.
Evidently, the entropy per base of a single strand is higher than that
of a double stranded DNA, and so a phase transition from the bound to
the unbound state is possible at $T_c= \epsilon/(2s_0-s_b)$.   This is
called the melting transition of a DNA.  A real melting is a slightly
more complicated phenomenon but this simple minded picture is
sufficient here for our purpose.  DNA strands can also be separated by
force at $T<T_c$ and that is called unzipping transition.

DNA strands need to be separated because, as per semi-conservative
replication, each of the new daughter molecules carry one of the
original strands.  DNA can be open or closed like a circular ring (or
a ribbon).  For example in many virus or bacteriophages, DNA is in an
open state but after infection it closes to form a circular DNA. 
A circular DNA (with $N\to\infty$) will also undergo similar melting
transition into two separate circular DNA as in Fig. \ref{fig:dna}b.
It is possible that there are other events disrupting a smooth joining
of the ends.  Suppose there is a $2\pi$ twist of the ribbon before
joining to form a twisted circular DNA (not a M\"obius strip).  This
long DNA will also undergo a melting transition to give two nonbonded
single stranded circles,  but the two circles are linked
topologically.   From a thermodynamics ot statistical mechanical point
of view, the slight change in the ``boundary conditions'' in the three
cases shown in Fig. \ref{fig:dna}  do not matter but from a biological
point of view, case (c) is dead or inactive because two the strands
cannot be shared by the daughter virus or phages.  We recognize the
importance of topological constraints in biology though it does not
affect the thermodynamical quantities like energy or entropy much.

\begin{figure}[htbp]
  \begin{center}
\includegraphics[scale=0.7,clip]{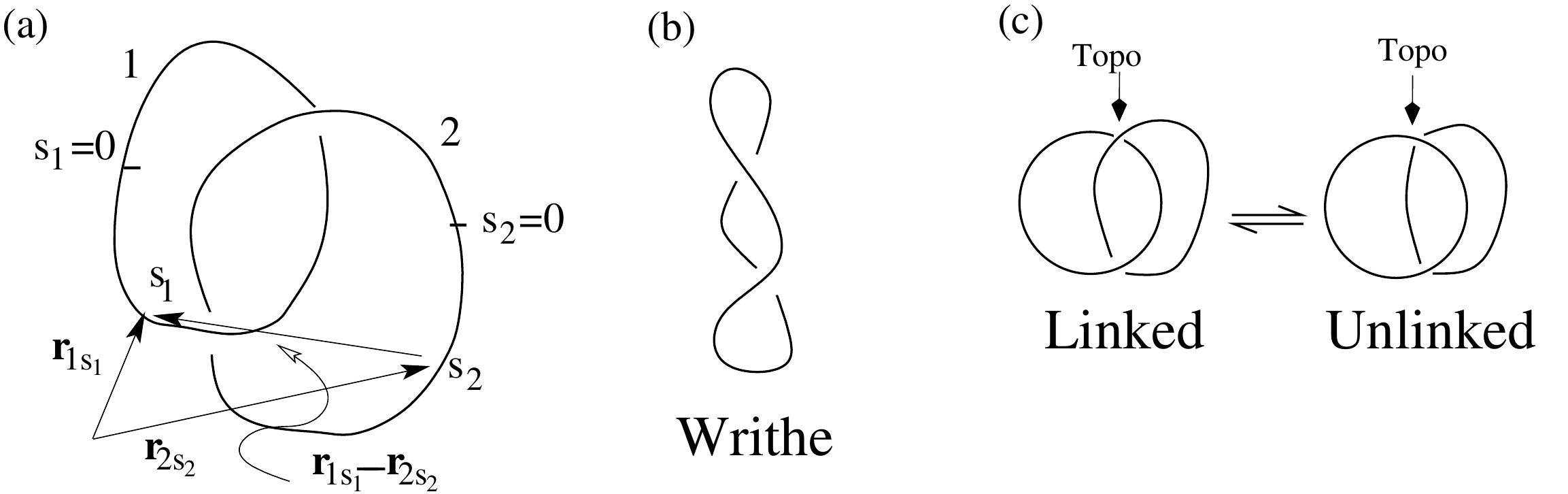}
\end{center}
  
\caption{(a) Two loops with contour variables $k_1$ and $k_2$, and
  position vectors used in Eq. (\ref{eq:2}). (b)The coiling of the
  axis of the loop is called writhe.  In DNA this leads to
  supercoiling.  (c) Two linked DNA with TopoIV (Topoisomerase IV is a
  type of Topoisomerase II).  TopoIV cuts both the starnds of a DNA,
  allows the other double stranded DNA to pass through the cut, and
  then rejoins the cut strands.  Left to right: TopoIV unlinks the two
  DNA's; right to left: two unlinked DNA's get topologically linked.}
  \label{fig:loop}
\end{figure}

\subsection{Linking number}
\label{sec:linking-number}

That cases (b) and (c) of Fig. \ref{fig:dna} are different can be seen by studying a
topological invariant called Gauss linking number.   For simplicity we
assume that each of the strand is free to cross itself (``phantom''
chain) but they are mutually avoiding.  In other words a chain cannot
cross the other.  This prevents the two circles becoming disentangled
in case (c).  If ${\bf r}_{is_i}$  denotes the
position vector of the point  of chain $i (i=1,2)$ at position
$s_i$ measured along the chain from an arbitrarily chosen point $s_i=0$,
then the Gauss linking number is given by\cite{mahan}
\begin{equation}
  \label{eq:2}
  Lk= \frac{1}{4\pi} \oint ds_1 \oint ds_2 \left[ 
             \left( \frac{\partial {\bf r}_{1s_1}}{\partial s_1}\right)\times  
             \left( \frac{\partial {\bf r}_{2s_2}}{\partial s_2}\right )\right ]\cdot 
\frac{{\bf r}_{1s_1}-{\bf r}_{2s_2}}{|{\bf r}_{1s_1}-{\bf r}_{2s_2}|^3},
\end{equation}
where the integrals are from $s_i=0,N$, assuming both the chains to be
of same length.  $Lk$ is zero for cases (a) and (b), but nonzero for
(c) for which it is 1.  Since each chain can cross itself, one of the
two, say 1, can be flattened into a planar circle.  The linking number
then counts the number of times  chain 2 pierces loop 1.  If
we put a direction on each of the two strands in the direction of
increasing $s_1,s_2$ as if there are currents in the loops, then the
counting  rule can be proved by using
Ampere's law in magnetism (see problem).  Because of the directions on
the loops, $Lk$ can be positive or negative.

The two integrals over the closed loops of the contour variables
$s_1,s_2$, can be rewritten as  a two dimensional integral over a
torus, ${\bf k}=(s_1,s_2)$, as 
$S^1\times S^1={\mathbb{T}}^2$.  If we define a unit vector
$${\hat{n}}({\bf k})= \frac{{\bf r}_{1s_1}-{\bf r}_{2s_2}}{|{\bf r}_{1s_1}-{\bf
    r}_{2s_2}|},$$ 
then the integral in Eq. (\ref{eq:2}) can be written
as
\begin{equation}
  \label{eq:28}
  Lk=\frac{1}{4\pi}\; \int\!\!\!\int_{\mathbb{T}^2} d^2 k \;\;  {\hat{n}}\cdot
\left(\frac{\partial {\hat{n}}}{\partial s_1}\times \frac{\partial
    {\hat{n}} }{\partial s_2}\right),
\end{equation}
which is identical in form as the Chern number integral for
insulators, Eq. (\ref{eq:27}).  Since ${\hat{n}}$ maps out a sphere, the
linking number actually counts how many times ${\hat{n}}$ winds around
the sphere as one covers the whole torus.  This is the same number one
looks for  Chern insulators, but now the context is vastly different.

\subsection{Twist and Writhe}
\label{sec:twist-writhe}

As mentioned already, the linking number remains invariant under
melting, which does not allow chain breaking.  For the unbound phase,
this is the only relevant topological invariant of interest, but for
the bound phase there can be other geometrical quantities.  Refer to
Fig. \ref{fig:bar}.  Consider a ribbon like object by making the
cross-section very thin.  The unit vector along the axis, ${\bf t}(s)$
represents the local DNA orientation in space, while ${\bf a}(s)$ is
the hydrogen bond vector at contour position $s$ measured along the
axis of the ribbon.  With the third direction, we have a triad, a
local coordinate system.  The ribbon or the bar is closed so that at
the end of the loop, ${\bf t}$ comes back to its original position
(${\bf t}(0)={\bf t}(L)$, thereby forming a closed loop on $S^2$ as in
Fig. \ref{fig:bar}c.  The solid angle formed by the loop is the change
in the orientation of ${\bf a}$.  The twist can be defined by
\begin{equation}
  \label{eq:30}
  Tw=\frac{1}{2\pi} \oint \; ds \; \; {\bf a}(s)\cdot  \left(\frac{\partial
      {{\bf a}}}{\partial s}\times \frac{\partial  {{\bf t}} }{\partial s}\right),
\end{equation}
which need not be an integer.  There is another geometric quantity
that determines the twisting of the axis as in Fig. \ref{fig:loop}b.
It is given by a formula similar to the linking number formula except
that both the integrations are over the same loop formed by the axis,
\begin{equation}
  \label{eq:31}
  Wr=\frac{1}{4\pi} \oint ds \oint ds' \left[ 
             \left( \frac{\partial {\bf r}_{s}}{\partial s}\right)\times  
             \left( \frac{\partial {\bf r}_{s'}}{\partial s'}\right )\right ]\cdot 
\frac{{\bf r}_{s}-{\bf r}_{s}}{|{\bf r}_{s}-{\bf r}_{s'}|^3}.
\end{equation}
Note that $s=s'$ does not pose a problem because the crossproduct is
also zero.   $Wr$ is a continuous variable and can be changed by
deforming the loop.  It depends only on the shape but not on the
scale.  The sign of $Wr$ tells us the overall handedness of the coil
-- right or left handed.
An important theorem by C{\u a}lug{\u a}reanu
connects the  topological invariant $Lk$ to the  two geometrical
quantities as
\begin{equation}
  \label{eq:32}
  Lk=Tw+Wr.
\end{equation}

If we now consider the ensemble of all possible closed configurations
of the double stranded DNA,  with the relaxed state (energetically minimum
state) with zero twist and zero writhe, we have $\langle
Lk\rangle=\langle Tw\rangle=\langle Wr\rangle =0$.  Intuitively, twist
and writhe are independent.  Therefore, the variances are additive, $\langle
Lk^2\rangle=\langle Tw^2\rangle+\langle Wr^2\rangle$. 
For a real DNA of length $L$ with a helical pitch of $\gamma$, the normal relaxed
state will have $Tw_0=L/\gamma$ In that case the averages satisfy
$\langle Tw-Tw_0\rangle=\langle Wr\rangle=0$, and $\langle (\Delta
Lk)^2\rangle=\langle (\Delta Tw)^2\rangle + \langle (\Delta
Wr)^2\rangle$ where $\Delta$ denotes deviation from the average.
The variances of
$Tw, Wr$ can be related to the elastic constants of the DNA, allowing
us to link topological or geometrical  features to the elastic constants.
Although,  $Lk$ remains constant at melting but $Tw, Wr$ lose
their meaning in the unbound phase, or even in partially unzipped state.

\subsection{Problem of Topoisomerase}
\label{sec:probl-topo}

Biological processes require trivial $L$.  Two closed loops with
different $L$ cannot be deformed into one another and therefore belong
to topologically different classes.  Note that this is true only in
three dimensions (${\bf r}$'s are 3-dimensional vectors).  An example
is shown in Appendix B on opening up  this link  in four
dimensions without any cut-paste.

There are enzymes called Topoisomerase II (topoisomerase IV to be
precise) that can cut a double stranded DNA at a crossing, change the
value of $Lk$ as shown in Fig. \ref{fig:loop}c, and rejoin the cut
DNA. This is needed in biology to separate two circular DNAs after
replication.  What is surprising is that the topoisomerase can locally
do a cut-paste to make the change.  In the figure only minimal number
of crossings are shown.  There could be many trivial crossings.  As
the topological feature is a global one (the integrations over the two
loops are equivalent to scanning the whole chains), how an object,
acting locally, can achieve this is a big puzzle.  It is easy to see
that, in thermal equilibrium, any crossing changed at random can
produce a link as often as it may open it up.  One wonders if
Topoisomerase knows of the fourth dimension!

\begin{problem}
Show the equality of the two expressions in Eqs. (\ref{eq:28}) and
(\ref{eq:2}).
\end{problem}
\begin{problem}
  Use the standard formula of magnetic field $d{\bf B}({\bf r})$ at
  ${\bf r}$ due to a small current element $d{\bf l}$ at ${\bf
    r}'(l)$, $d{\bf B}({\bf r}) \propto d{\bf l}\times {\bf\nabla}
  \frac{1}{|{\bf r}-{\bf r}'(l)|}$ and Stokes' theorem to prove that
  $Lk$ in Eq. (\ref{eq:2}) is an integer..
\end{problem}
\begin{problem}
Show that $Tw$ is additive, i.e. it can be computed by adding twists
for  pieces.  Show that writhe is zero for a planar figure.  
\end{problem}
\begin{problem}
Prove C{\u{a}}lug{\u{a}}reanu's theorem.
\end{problem}
\begin{problem}
Show that for a  double starnded  DNA loop, Topoisomerase IV
changes writhe by $\pm 2$.
\end{problem}

\section{Summary}
\label{sec:summary}
In this chapter we explored several simple problems from classical
mechanics, statistical mechanics, and quantum mechanics by using
topological arguments.  These, in turn, allowed us to explain some of
the basic ideas of elementary topology in terms of the known physical
phenomena, and the common link among diverse topics.
\pagebreak

\section*{Appendix A: Mobius strip and Stokes' theorem}
\label{sec:appendix-a:-mobius}
\addcontentsline{toc}{section}{Appendix A: M\"obius strip and Stokes
  theorem}

Let us now discuss a different type of problems involving integrals of
vector fields.  Stokes' theorem states that the surface integral of
the curl of a vector field is equal to the line integral of the field
over the boundary of the surface.
 
\newcommand{\figmob}{%
 \begin{figure}[htbp]
   \begin{center}
       \includegraphics[scale=0.5,clip]{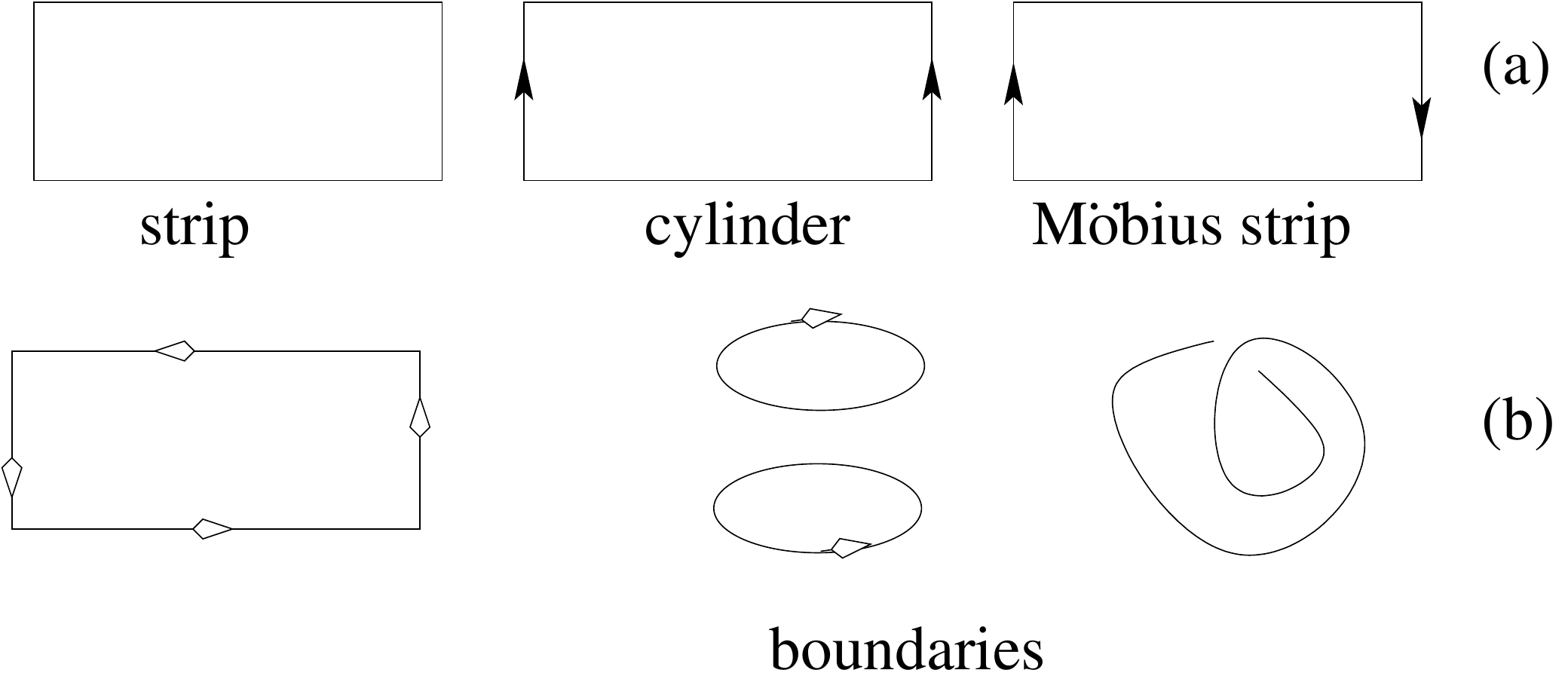}
    \end{center}
 \caption{Cylinder, M\"obius strip and Stoke's law.}\label{fig:mob}
\end{figure}
 }%

 \figmob

 Take a rectangular strip (say, a long piece of paper) of length $2\pi
 {{R}}$ and width $2L$ as in Fig. \ref{fig:mob}a.  Now use the arrows
 to define ``equivalence'' condition, namely periodic boundary
 condition in the x-direction to get a cylinder and with a twisted
 boundary condition (``half-twist'') to get a M\"obius strip.  These
 two in 3 dimensions can be described by
 \begin{eqnarray}
   \label{eq:2app}
   {\rm cylinder:}&& x(t,\theta)=R \cos\theta, y(t,\theta)=R \sin\theta,
   z(t,\theta)=t,\\
 {\rm Mobius:}&& x(t,\theta)=\left(R-t \sin\frac{\theta}{2}\right)
 \cos\theta, y(t,\theta)=\left(R-t \sin\frac{\theta}{2}\right)
 \sin\theta, \nonumber\\
 &&  z(t,\theta)=t \cos\theta,\label{eq:5app}
 \end{eqnarray}
 for $-L\leq t\leq L$ and $0\leq\theta<2\pi$.

 The cylinder has two boundaries at $t=-L$ and $t=L$, but the M\"obius
 strip has only one boundary.  For the M\"obius strip, we reach the
 same point by going around twice, so that the boundary can be
 described by Eq. (\ref{eq:5app}) with $t=L$ and $0\leq \theta\leq 4\pi$.
 It actually consists of the two original boundaries at $t=L, 0\leq \theta\leq 2\pi$
 and  $t=-L, 0\leq \theta\leq 2\pi$ joined together
 to form an unknot, as can be seen in Fig.
 \ref{fig:mob}b.

 Suppose we have a vector field 
$${\bf A}=\hat{k}\times \frac{{\vec{\rho}}}{\rho^2},$$ 
where $\vec{ \rho}=x \hat{\imath}+y\hat{\jmath}$ in the same three
dimensional coordinate system.  For this field $\nabla\times {\bf
  A}=0$ everywhere except for the z-axis.

We now consider the three geometries of Fig.  \ref{fig:mob}a
separately.
\begin{enumerate}
\item  Let us now put the strip in Fig. \ref{fig:mob} say parallel to the z
 axis with the centre at $(R,0,0)$.  The surface integral $\int\!\int
 d{\vec S}\cdot (\nabla\times{\bf A}) =0$.  It is easy to check that 
 Stokes' theorem is valid by showing that the line integral $\oint {\bf
 A}\cdot d{\bf l}=0$, where the integration is along the boundary, as shown
in the left figure of  Fig.  \ref{fig:mob}b.

\item For the cylinder placed along the z-axis with the centre of the
  cylinder at the origin, there are two boundaries along which one has
  to do the line integral.  Since the directions are opposite for the
  two rings at $z=\pm L$ (middle fig of Fig.  \ref{fig:mob}b), the
  total line integral is zero.  In other words  Stokes' theorem is
  explicitly verified.

\item  For the M\"obius case, the line integral along the boundary curve  is
 $$\oint_{\rm boundary} {\bf A}\cdot d{\bf r}=\int_0^{4\pi}  {\bf
   A}(x(\theta),y(\theta),z(\theta))\cdot {\bf r}^{\prime}\
 {d\theta}\neq 0,$$
where  ${\bf r}^{\prime}=d {\bf r}/d\theta$.  For the parametrization
 used, ${\bf A}\cdot{\bf r}^{\prime}=1$ so that the integral is
 $4\pi$.  {\it There seems to be a violation of Stokes' law.}
\end{enumerate}

This paradox is resolved by noting that the area vector $d{\bf S}$
cannot be defined on the M\"obius strip.  If we slide a small area
element along the strip through $2\pi$ the area vector will not point
in the same direction.  Another way of seeing the difference is to
colour the surfaces without any abrupt change.  Two colours are needed
to paint the surfaces of the strp and the cylinder, but one is enough
for the M\"obius strip.  Such a surface, like the M\"obius strip, is
called a nonorientable surface and Stokes' theorem is not applicable
there.


 \begin{problem}
 Is it possible to generalize the M\"obius strip construction so that
 the boundary curve is, say, a trefoil knot? (Hint: three half-twists)
 \end{problem}

\section*{Appendix B: Disentanglement via moves in 4-dimensions}
\label{sec:appendix:-moves-4}
\addcontentsline{toc}{section}{Appendix B: Disentanglement via moves in
  4-dimensions}

\begin{figure}[htbp]
  \begin{center}
\includegraphics[scale=0.5]{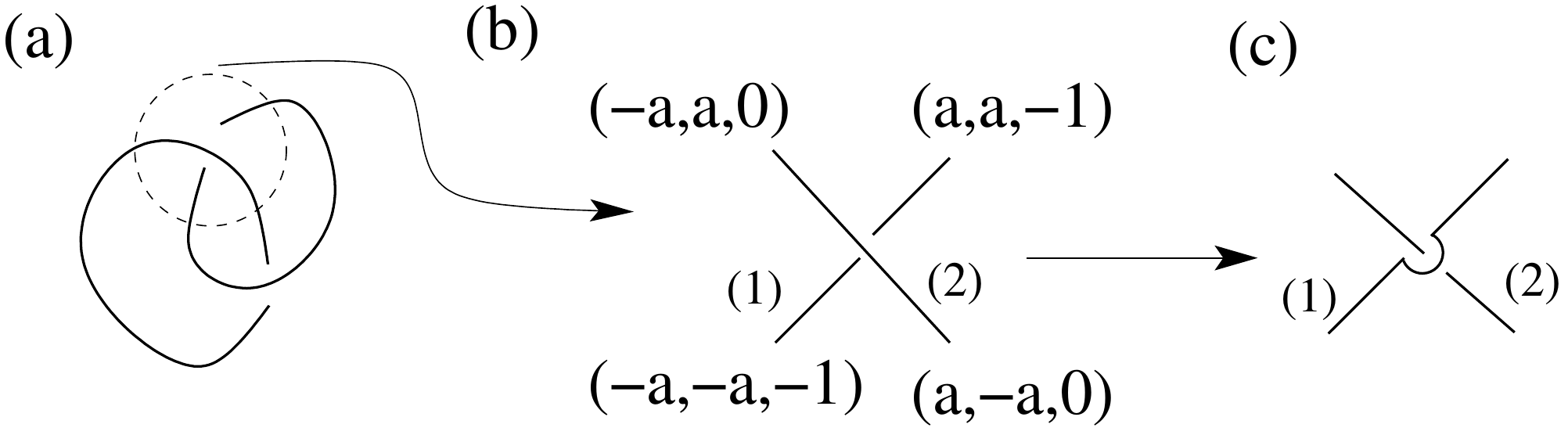}
\end{center}
\caption{(a) Linked loops.  The crossing indicted by the dotted circle
  is shown in (b).  In three dimensions, chain 1 in (b) is in the
  plane $z=-1$ while the other one is at $z=0$ so that chain 1 is
  below chain 2.  The crossing point is taken to be at $x=y=0$ when
  projected in the xy plane.  If the chain configurations can be
  changed to type (c), then the two loops can be unlinked. This is
  possible in 4-dimensions but not in three.}
  \label{fig:cross}
\end{figure}

Take two loops with linking number one.  It is a common knowledge that
the two loops cannot be taken apart if the chains do  not cross each
other.  We now show a set of local moves in four dimensions (x-y-z-w space) that takes
(b) to (c) in Fig. \ref{fig:cross}.\footnote{This is based on the online mathjournal article by Jeff Boersema and Erica J. Taylor,
https://www.rose-hulman.edu/mathjournal/archives/2003/vol4-n2/paper2/v4n2-2pd.pdf}

Let $t$ be a parameter, $t\in [0,1]$, so that chain 1 can be
parameterized in 4-dimensions by 
\begin{eqnarray}
  \label{eq:8}
  f_0(t)=\left\{\begin{array}{ll}
(-a,-a,-{1},0)&\quad (0\leq t\leq \frac{3}{8})\\[5pt]
(-a+8a(t-\frac{1}{8}),-a+8a(t-\frac{1}{8}),-1,0)&\quad (\frac{3}{8}\leq t\leq \frac{5}{8})\\[5pt]
(a,a,-1,0)&\quad (\frac{5}{8}\leq t\leq 1)
\end{array}\right.
\end{eqnarray}
We  shall deform the chain in three mutually exclusive  ways, (i) in the
xy plane, or, 
(ii) in the z-direction, or, (iii) in the 4th w direction.   The
variable $t$ is used to describe the configuration of the chain, while
a second variable $s \in [0,1]$ is to be used for the deformation so
that the chain is described as $f_s(t)$. 

First we deform in the x-y-w space keeping $z=-1$ fixed.
The new chain at $s=1/3$  is given by
 \begin{eqnarray}
   \label{eq:9}
   f_{1/3}(t)=\left\{  \begin{array}{ll}
 (-a,-a,-1,8t),&\quad  (0\leq t\leq\frac{1}{8})\\[5pt]
 (-a,-a,-1,1),&\quad  (\frac{1}{8}\leq t\leq\frac{3}{8})\\[5pt]
 (-a+8a(t-\frac{3}{8}),-a+8a(t-\frac{3}{8}),-{1},1),&\quad  (\frac{3}{8}\leq t\leq\frac{5}{8})\\[5pt]
 (a,a,-1,1),&\quad  (\frac{5}{8}\leq t\leq\frac{7}{8})\\[5pt]
 (a,a,-1,1-8(t-\frac{7}{8})),&\quad
 (\frac{7}{8}\leq t\leq 1)
  \end{array}\right.
 \end{eqnarray}
Next, we deform the $z$ coordinate from -1 to 1 taking the chain above
the other one as
\begin{eqnarray}
  \label{eq:10}
    f_{2/3}(t)=\left\{  \begin{array}{ll}
 (-a,-a,-1,8t),&\quad  (0\leq t\leq\frac{1}{8})\\[5pt]
 (-a,-a,-1+16(t-\frac{1}{8}),1),&\quad  (\frac{1}{8}\leq t\leq\frac{2}{8})\\[5pt]
 (-a,-a,1,1),&\quad  (\frac{2}{8}\leq t\leq\frac{3}{8})\\[5pt]
 (-a+8a(t-\frac{3}{8}),-a+8a(t-\frac{3}{8}),{1},1),&\quad  (\frac{3}{8}\leq t\leq\frac{5}{8})\\[5pt]
 (a,a,1,1),&\quad  (\frac{5}{8}\leq t\leq\frac{6}{8})\\[6pt]
 (a,a,1-16(t-\frac{6}{8}),1),&\quad  (\frac{5}{8}\leq t\leq\frac{7}{8})\\[6pt]
 (a,a,-1,1-8(t-\frac{7}{8})),&\quad
 (\frac{7}{8}\leq t\leq 1)
  \end{array}\right. 
\end{eqnarray}
Finally, we bring back the fourth $w$ coordinate to zero as,
\begin{eqnarray}
  \label{eq:11}
    f_{1}(t)=\left\{  \begin{array}{ll}
 (-a,-a,-1,0),&\quad  (0\leq t\leq\frac{1}{8})\\[5pt]
 (-a,-a,-1+16(t-\frac{1}{8}),0),&\quad  (\frac{1}{8}\leq t\leq\frac{2}{8})\\[5pt]
 (-a,-a,1,0),&\quad  (\frac{2}{8}\leq t\leq\frac{3}{8})\\[5pt]
 (-a+8a(t-\frac{3}{8}),-a+8a(t-\frac{3}{8}),{1},0),&\quad  (\frac{3}{8}\leq t\leq\frac{5}{8})\\[5pt]
 (a,a,1,0),&\quad  (\frac{5}{8}\leq t\leq\frac{6}{8})\\[6pt]
 (a,a,1-16(t-\frac{6}{8}),0),&\quad  (\frac{5}{8}\leq t\leq\frac{7}{8})\\[6pt]
 (a,a,-1,0),&\quad
 (\frac{7}{8}\leq t\leq 1)
  \end{array}\right.   
\end{eqnarray}
It is now straight forward to construct a map $f_s(t)$ linear and
continuous in $s$ -- it is already linear and continuous in $t$ --
that goes from $f_0(t)\to f_{1/3}(t) \to f_{2/3}(t) \to f_1(t)$.  For
example, for $2/3\leq s\leq 1$ a continuous map from $f_{2/3}(t)$ to
$f_1(t)$ can be constructed by replacing the $w$ values of $f_{2/3}$ by
$3w(1-s)$. Note that we get back the same boundary points
$(-a,-a,-1,0)$ at $t=0$ and $(a,a,-1,0)$ at $t=1$ for all $s\in
[2/3,1]$.

The two loops can  therefore be delinked via these local  deformations in the 4th dimension.


\begin{thebibliography}{99}
\bibitem{naka} M. Nakahara, {\it Geometry, Topology and Physics}
  (Taylor \& Francis, Boca Raton, FL, USA, 2003).
\bibitem{ald}  R. Aldrovandi and  J. G. Pereira, {\it An Introduction
    to Geometrical Physics} (World Scientific, SIngapore, 1995).
\bibitem{nash} C. Nash, S. Sen, {\it Topology and Geometry for
    Physicists} (Dover, Mineola, NY, 2011).
\bibitem{somnath}  See, e.g.,  the chapters by Somnath Basu, and by
  Atreyee Bhattacharya in this book.
\bibitem{samik}  See, e.g., the chapters by Samik Basu, and the
  tutorial by Soma Maity  in this book.
\bibitem{dheeraj}  See, e.g., the chapter by Dheeraj Kulkarni  in this book.
\bibitem{wilc}  {\it Geometric Phases in Physics},
Edited by: F Wilczek and  A Shapere (World Scientific, Singapore, 1989)
\bibitem{mahan}  See, e.g., the chapters by Mahan Mj, and by
  P. Ramadevi  in this book.

\bibitem{pt}  For an example of an insulator with Chern number=2, see
P. Titum, N. H. Lindner, M. C. Rechtsman, and G. Refael, 
Phys. Rev. Lett. {\bf 114}, 056801 (2015).
\end{thebibliography}
\end{document}